# The Nature of High-frequency Oscillations Associated with Short-lived Spicule-type Events

Juie Shetye[1,2], Erwin Verwichte[3], Marco Stangalini[4], and J. G. Doyle[2]  
[1] Dept. of Astronomy, New Mexico State University, Las Cruces, NM, USA; jshetye@nmsu.edu  
[2] Armagh Observatory and Planetarium, College Hill, Armagh BT61 9DG, UK  
[3] Centre for Fusion, Space and Astrophysics, University of Warwick, Coventry CV4 7AL, UK  
[4] ASI—Italian Space Agency, Via del Politecnico snc, 00133, Rome, Italy

## Abstract

We investigate high-resolution spectroscopic and imaging observations from the CRisp Imaging SpectroPolarimeter (CRISP) instrument to study the dynamics of chromospheric spicule-type events. It is widely accepted that chromospheric fine structures are waveguides for several types of magnetohydrodynamic (MHD) oscillations, which can transport energy from the lower to upper layers of the Sun. We provide a statistical study of 30 high-frequency waves associated with spicule-type events. These high-frequency oscillations have two components of transverse motions: the plane-of-sky (POS) motion and the line-of-sight (LOS) motion. We focus on single isolated spicules and track the POS using time–distance analysis and in the LOS direction using Doppler information. We use moment analysis to find the relation between the two motions. The composition of these two motions suggests that the wave has a helical structure. The oscillations do not have phase differences between points along the structure. This may be the result of the oscillation being a standing mode, or that propagation is mostly in the perpendicular direction. There is evidence of fast magnetoacoustic wave fronts propagating across these structures. To conclude, we hypothesize that the compression and rarefaction of passing magnetoacoustic waves may influence the appearance of spicule-type events, not only by contributing to moving them in and out of the wing of the spectral line but also through the creation of density enhancements and an increase in opacity in the H$\alpha$ line.

*Unified Astronomy Thesaurus concepts:* Solar chromosphere (1479); Solar atmosphere (1477); Magnetohydrodynamics (1964)

## 1. Introduction

Solar spicules are dynamic jet-like structures observed at chromospheric temperatures that typically live for a few minutes and reach heights of 3000–4000 km in the solar atmosphere (Thomas 1948; Beckers 1968, 1972; Sterling 1998, and references therein). Advances in ground-based instrumentation, such as CRisp Imaging SpectroPolarimeter (CRISP; Scharmer et al. 2008) on the Swedish 1 m Solar Telescope (SST; Scharmer et al. 2003) and the Interferometric BIdimensional Spectropolarimeter (IBIS; Cavallini 2006) on the Dunn Solar Telescope (DST; Dunn 1969), have made it possible to investigate spicules in unprecedented detail. This has revealed various types of spicules: the classical type I spicule (Thomas 1948; Beckers 1968, 1972; Sterling 1998); a new, faster-evolving spicule entitled type II (De Pontieu et al. 2007b); plus an on-disk counterpart. These are subdivided into two categories with a duration up to a few minutes: rapid blueshifted excursions (RBEs; Langangen et al. 2008; Rouppe van der Voort et al. 2009; Tsiropoula et al. 2012; Sekse et al. 2013a) and rapid redshifted excursions (RREs; Sekse et al. 2013b). More recently, high-cadence data have revealed a new type of spicule-type event that suddenly appears in the field of view (FOV) of the narrowband filter without any evidence of evolution (Judge et al. 2011, 2012; Lipartito et al. 2014; Shetye et al. 2016b; Pereira et al. 2016).

Spicules are thought to support several types of waves, which have been associated with spicule formation (Uchida 1961; Hollweg et al. 1982; Suematsu et al. 1982; Shibata & Suematsu 1982; Haerendel 1992; De Pontieu & Haerendel 1998; De Pontieu 1999; Sterling 2000; James et al. 2003; De Pontieu et al. 2004) and energy deposition (Nikolsky & Platova 1971; Kukhianidze et al. 2006; Zaqarashvili et al. 2007; De Pontieu et al. 2007a; He et al. 2009a, 2009b; Zaqarashvili & Erdélyi 2009; Okamoto & De Pontieu 2011; De Pontieu et al. 2012; Kuridze et al. 2012, 2013; Morton et al. 2012; De Pontieu et al. 2014a; Kuridze et al. 2015; Henriques et al. 2016; Shetye et al. 2016a, henceforth JS16). The most prominent wave signatures detected are periodic transverse displacements of the spicule axis. They have been interpreted as fast kink MHD waves, i.e., Alfvénic waves (see review by Zaqarashvili & Erdélyi 2009). However, the presence of standing waves has been reported by Okamoto & De Pontieu (2011). Furthermore, waves along spicules are suspected to accelerate the solar wind (De Pontieu et al. 2007a, and references therein), making spicules extensively studied events in the solar atmosphere.

Shetye et al. (2016b) and JS16 investigated the properties and dynamics of a set of spicule-type structures or events. These events characteristically appear and disappear suddenly in narrowband spectral images. This has been attributed to mass motions (Pereira et al. 2016) or sheets of chromospheric material (Judge et al. 2011, 2012; Lipartito et al. 2014). Shetye et al. (2016b) further reported on transverse oscillations supported by these spicule-type events. They typically have displacement amplitudes of 230 km, periods of 20 s, and velocity amplitudes of 15–30 km s$^{-1}$. The importance of analyzing such high-frequency motions associated with spicule-type events arises from the theory or idea that these motions can reach the corona (Srivastava et al. 2017). The findings in the present manuscript are focused on a relatively untouched aspect of wave propagation that is associated with a number of spicule-type events. The analysis of a statistically significant number of spicules is required to assess the typical range of oscillations occurring in spicules. In this study we extend the sample to 30 new events that exhibit oscillatory motions that have frequencies in the relatively unexplored range of 10–50 mHz, which is higher than the dominant natural frequency of 5 mHz found in the chromosphere (Fleck & Schmitz 1991).

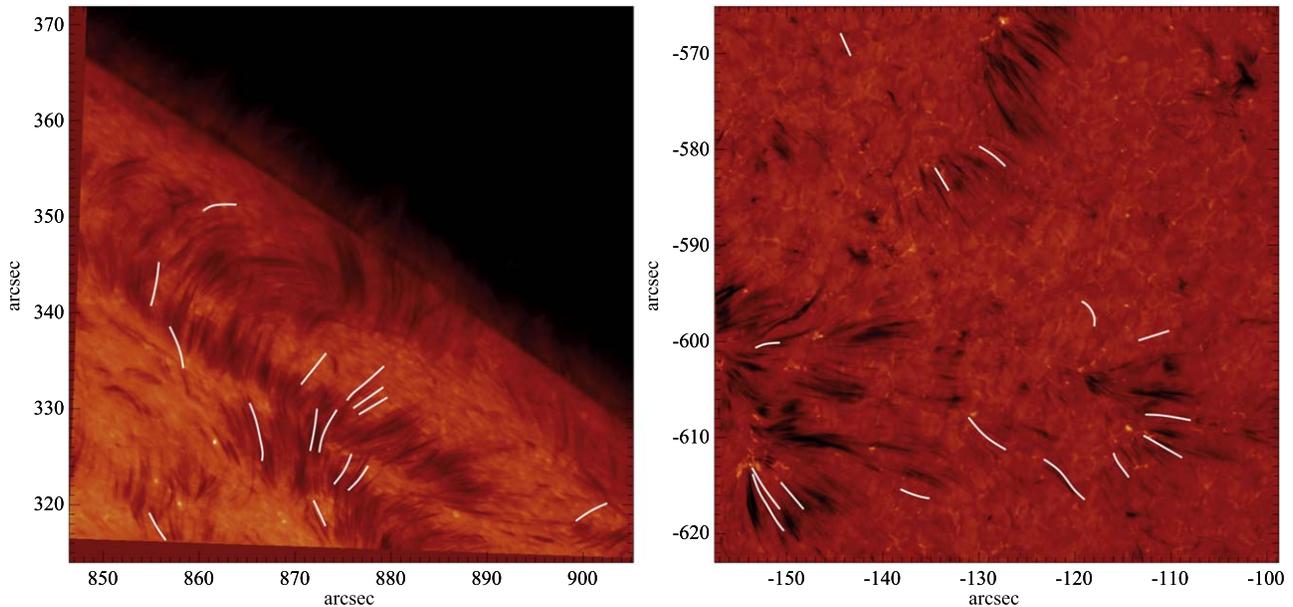

**Figure 1.** Context images obtained at −774 mÅ, showing the location of spicule-type events (solid white lines) in both data sets. Left: from a limb data set obtained on 2014 June 5. Right: from a coronal hole data set obtained on 2014 June 7.

For longer-lasting and relatively slower evolving spicules, oscillatory behavior has been identified in line-of-sight (LOS) and plane-of-sky (POS) displacements (Sharma et al. 2017, 2018). The identification of the nature and polarization of these oscillations requires the combination of the wave signatures from both components of transverse movement (Zaqarashvili & Skhirtladze 2008; Sharma et al. 2018). Our motivation is to focus on the high-frequency motion, where such insight into the POS and LOS motions would help us understand the nature of the wave present and shed light onto how the wave might propagate.

The paper is organized as follows. Section 2 provides the basic information of the observational data sets used. Section 3 outlines the methodology. In Section 4, several case studies are presented. This includes an investigation of the phase relation between LOS and POS signals where they are both present. In Section 5, the statistical properties of spicule-type events are shown. Finally, the discussion and conclusions are summarized in Section 6.

## 2. Observations

We analyze spectral imaging observations obtained by the CRISP (Scharmer et al. 2008) instrument on the SST (Scharmer et al. 2003). CRISP has an FOV of $60'' \times 60''$ with an image scale of $0\rlap{.}''0592$. The telescope uses an adaptive optic system consisting of a tip-tilt mirror and an 85-electrode deformable mirror setup to assist our observations. The diffraction-limited resolution at H$\alpha$ is $0\rlap{.}''059$. The H$\alpha$ prefilter FWHM is 4.9 Å, while the FWHM of the transmission filter is 60.4 mÅ, allowing observations at multiple positions within the H$\alpha$ line profile. After Multi-object Multi-frame Blind Deconvolution (MOMFBD; van Noort et al. 2005) reconstruction, an effective science-ready cadence between 4 and 5 s was achieved. Figure 1 shows the two data sets.

### 2.1. 2014 June 5

Data set 1 is a sequence of solar limb observations on 2014 June 5 from 11:53 UT to 12:34 UT in H$\alpha$ and Ca II $\lambda$8542. The CRISP instrument was pointed at $x_c = 876''$, $y_c = 343''$. The H$\alpha$ $\lambda$6563 sequence consists of nine spectral line positions spaced with an increment of 258 mÅ from the line center in the red and blue wings. Ca II $\lambda$8542 has nine line positions incremented at $\pm 55$ Å from the line core. Eight frames were taken at each of the above 18 wavelength positions, resulting in a cadence of 5 s. This observation sequence has previously been studied by Shetye et al. (2016b).

### 2.2. 2014 June 7

Data set 2 is a sequence obtained on 2014 June 7 from 07:21 UT to 09:21 UT with CRISP FOV centered at $x_c = 128''$, $y_c = -594''$, on the boundary of a low-latitude coronal hole covering a quiet-Sun region. The observations in the H$\alpha$ spectral line positions were obtained at seven line positions centered at 6563.0 Å, $\pm 260$, $\pm 774$, and $\pm 1032$ mÅ, and the Ca II $\lambda$8542 line at seven wavelength positions (i.e., the line center 8542 Å $\pm 55$, $\pm 110$, and $\pm 495$ mÅ) using a narrowband filter of 0.111 Å.

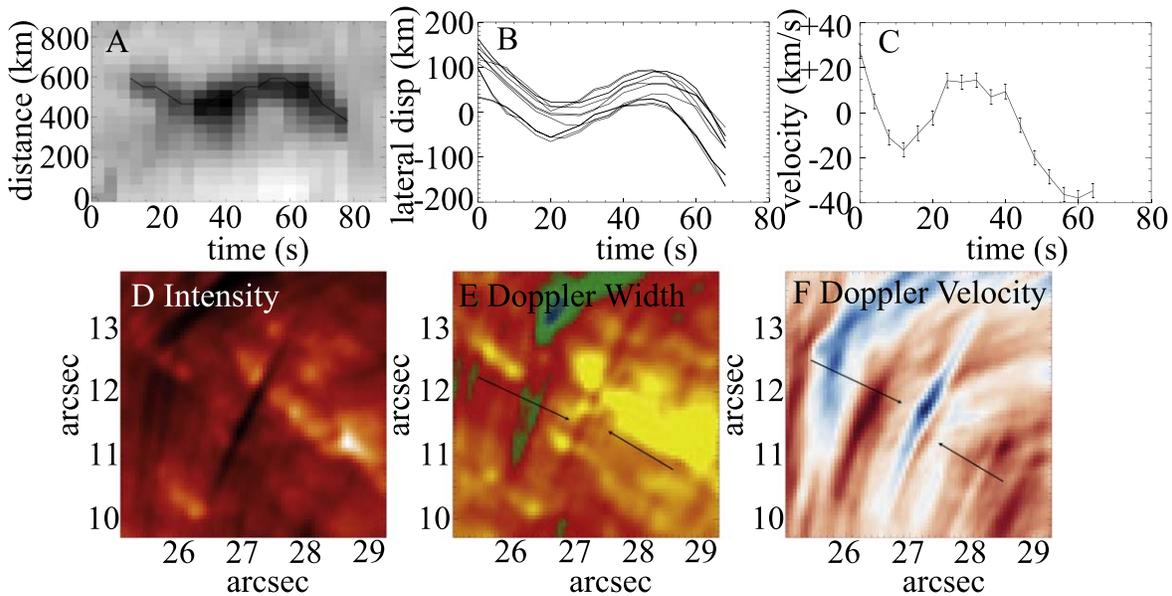

**Figure 2.** The top row shows a sample tracked path of the spicule-type event: (a) time–distance plot; (b) POS transverse motion traced using a Fourier phase method, for 10 slit positions along the length of spicule; (c) image-plane transverse velocity. The error bars show standard errors. The bottom row shows results of a moment analysis: (d) intensity image; (e) results from the Doppler width calculation; (f) results from the Doppler velocity. Black arrows on panels (e) and (f) indicate the location of the structures.

### 3. Methodology

We identify by eye spicule-type events with interesting dynamics. The length of a spicule is defined as the maximum POS projected extent. The duration of the time series is limited to the lifetime of the event, which is obtained at the best observed wavelength position in H$\alpha$. All events in this study undergo transverse motions in the POS and LOS directions. We follow the method set out in Shetye et al. (2016b) to compute time series of the spicule's lateral position at various locations along the spicule. The method is summarized as follows. From an image cut across the structure, the position of the local minimum in the intensity, $x(t)$, is extracted at each time at a subpixel accuracy using a Fourier phase correlation method (Mohamed et al. 2012). The average value of the position of the structure $\bar{x}(t)$ is subtracted from $x(t)$ to separate the lateral displacement, $\delta x(t)$, from bulk motion, $\bar{x}(t)$, at any given time, i.e., $x(t) = \bar{x}(t) + \delta x(t)$. Figure 2 shows an example of tracking performed for a spicule-type event. In the POS we measure the oscillation amplitude as the maximum transverse displacement of the spicule-type event. Where more than one oscillation cycle is visible, the oscillation period is taken as the average time interval between subsequent maxima (or minima) in the displacement time series. A similar approach was used in Srivastava et al. (2017). An estimate of the error in the spicule-type events' lateral position is found from the standard deviation of the absolute values of all the lateral displacements during the event, i.e., SD($|\delta x(t)|$). The standard deviation would also be affected by intrinsic dynamics of the spicule, such as perturbations along the motion of the spicule.

Figure 3 shows the POS transverse motion for the spicule-type event #5 from Table A1. This event is particularly difficult to trace owing to the vicinity of other structures in the LOS. We plot 10 slits along the length of the spicule. Panels (d)–(f) show the POS transverse motion with varying amplitude and period. For such events we note that the POS motion has a decaying amplitude and period. Due to the short lifetime of the spicule, we are not able to comment on the harmonic nature of the oscillation.

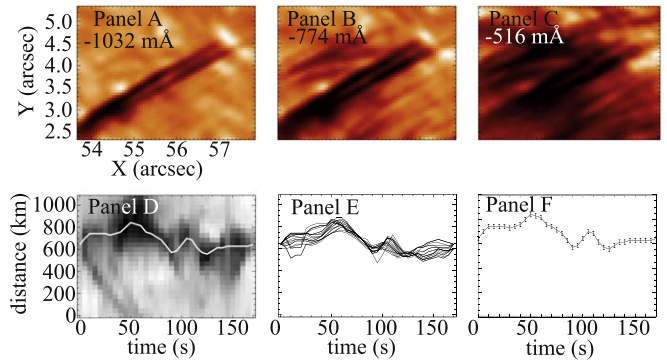

**Figure 3.** The top row shows the time stamp of the spicule-type event #5 from Table A1, observed at −1032, −774, and 516 mÅ, left to right. The bottom row shows the POS transverse motion. (d) Time–distance plot with tracking overplotted in white solid line. (e) Time–distance plot with tracking shown for 20 slits along the length of the event. (f) The time–distance slit with standard error.

We also see in the time–distance plot of Figures 2 and 3 the presence of nearly parallel linear tracks that repeat in synchronization with displacements toward smaller distance values. These tracks are suggestive of passing wave fronts. The obvious candidate is the fast magnetoacoustic wave, which is a nearly isotropic wave that is able to propagate across magnetic field lines. We estimate from the time–distance plot that these waves propagate at a POS projected speed of 13–18 km s$^{-1}$.

We use a moment analysis on the H$\alpha$ line scans to compute intensity, Doppler velocity, and Doppler width corresponding to a spicule-type event similar to Rouppe van der Voort et al. (2009). The intensity at each pixel is assumed to be a function of wavelength $\lambda$ within the range of a particular spectral line at rest wavelength $\lambda_0$, i.e., $I(\lambda)$. The $p$th moment of this spectral

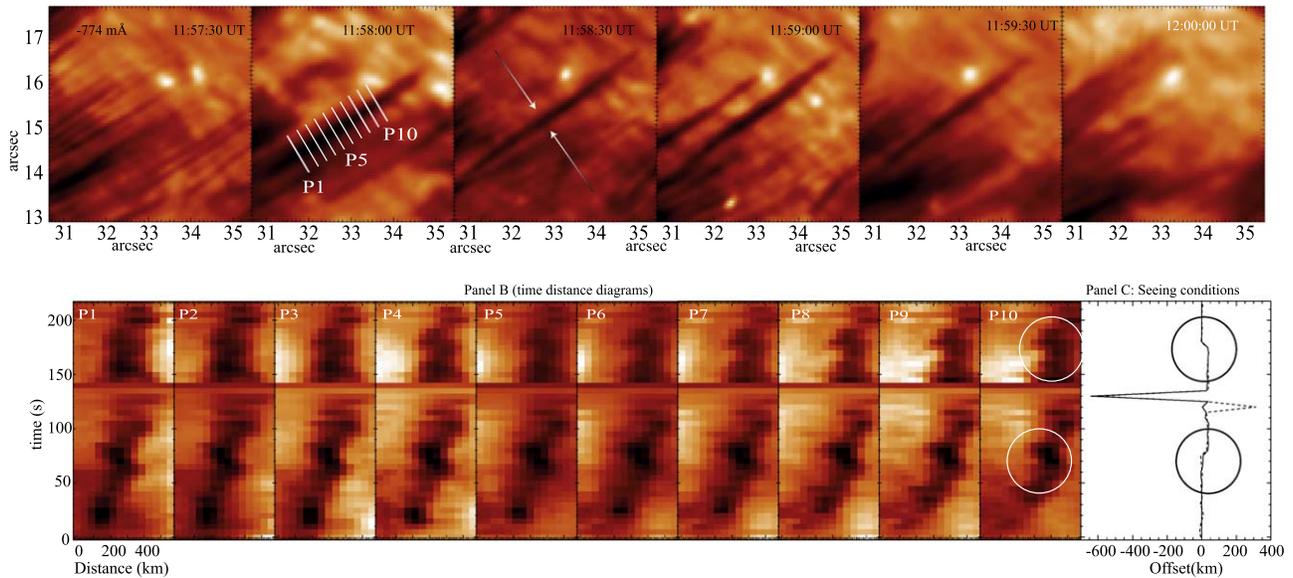

**Figure 4.** (a) Evolution of a spicule-type event between 11:57:30 UT and 12:00:00 UT at −774 mÅ. White arrows show the location of the event. The second subpanel shows the location of time–distance slits P1−P10. (c) Spicule-type event's motion observed in dark. (d) Image offset plot throughout the lifetime of the spicule, showing a shift in x (solid) and y (dashed) directions.

line is defined as

$$m_p = \int_0^\infty I(\lambda)(\lambda - \lambda_0)^p d\lambda. \quad (1)$$

The moments are related to the total intensity, $\bar{I} = m_0$, Doppler velocity, $v_D$, and Doppler width, $\Delta v_D$, respectively, as

$$\bar{I} = m_0, \quad v_D = \frac{c}{\lambda_0}\frac{m_1}{m_0}, \quad \Delta v_D = \sqrt{\frac{c}{\lambda_0}\frac{m_2}{m_0}}, \quad (2)$$

where $c$ is the speed of light in a vacuum. These spectral quantities are computed at each time and at each pixel along an image cut following the minima tracked during the POS motion of the spicule-type event. The resulting time series represents the LOS transverse motion of the event. The procedure is repeated for various locations along the spicule axis. Figure 2 shows the result of moment analysis for a test example.

We explore the relationship between the two orthogonal components of transverse motion for events where we detect oscillations in both the POS and LOS time series. The phase relation between the two components provides insight into the polarization of the oscillation mode. Furthermore, a comparison of the phase at different locations along the structure can provide insight into the propagation of the wave mode. We investigate the phase relation between points at the two ends and halfway along the spicule.

We identify the dominant periodicities in the two time series $t_1$ and $t_2$ using the continuous wavelet transform (CWT; Torrence & Compo 1998) using a Morlet mother wavelet. The cross-correlation (XWT) of the two wavelet transforms allows the examination of the phase relations between the POS and LOS time series:

$$\text{XWT}(t_1, t_2) = \text{CWT}(t_1)\text{CWT}(t_2)^*, \quad (3)$$

where $^*$ indicates the complex conjugate. The XWT reveals areas in time period parameter space with common high power. Where the cross-correlation is statistically significant in the XWT, we use arrows to clarify the nature of the correlation. The length of the arrow is indicative of the correlation strength. The angle of the arrow reveals the phase relation. Where both time series are in phase (antiphase), i.e., the phase angle is 0° (180°), the arrows point to the right (left). For a time series t1 that leads (lags) t2 by 90°, the vector points down (up). Furthermore, the correlation strength is shown by contours in the wavelet plots of Figures 9, 13, 11, and 15.

### 3.1. Post-destretching Correction in the Co-alignment

To check if spicule displacements are solar in origin and not associated with post-alignment errors, we track residual image offsets or jitter between subsequent images in both x- and y-directions using shift correlation applied to the sequence of images where spicule-type events are observed. The resulting time series are shown next to all event time–distance plots. For all events the residual image jitter is significantly lower than the displacement amplitude of oscillations. However, large deviations are expected when seeing deteriorates, although seeing aberrations have much shorter timescales of the order of tens of milliseconds, thus not consistent with the detected periodicities (Stangalini et al. 2017). We illustrate this in Figure 4, which shows the evolution of an event observed in the blue-wing position −774 mÅ observed between 11:57 and 12:00 UT in data set 1. The time–distance plot at 10 positions shows a structure with significant and long-lasting oscillation. During this time, the seeing deteriorates and is noticeable as a discontinuity in the time–distance image (see panel (b)). The evolution of image offset between two subsequent images shows a large deviation (see panel (c)). The motion of the spicule-type event observed during this period also shows a region highlighted by circles where the offset becomes significant. There are also smaller deviations (highlighted by black circles) that indicate a temporary horizontal shift around the time of deteriorated seeing.

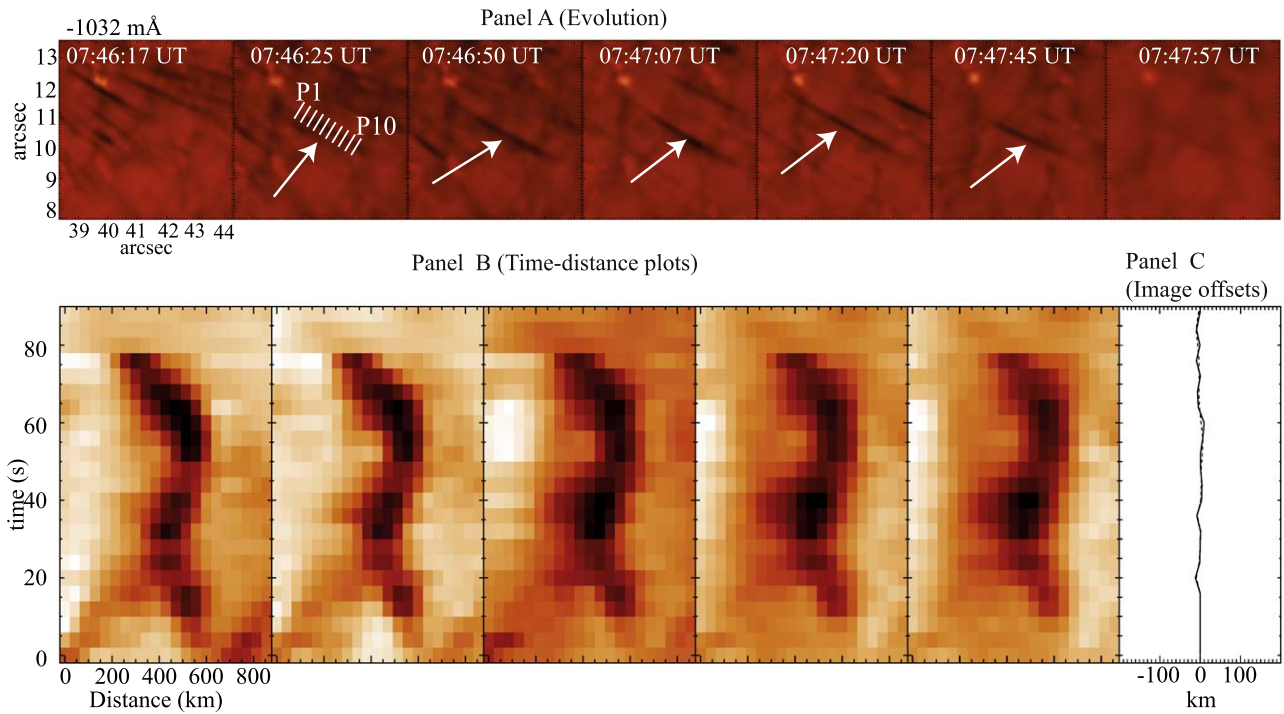

**Figure 5.** (a) Evolution of a spicule-type event #20 from Table A3 and shown in Figure B3, observed between 07:46:50 UT and 07:47:45 UT at −1032 mÅ in Hα. White arrows show the location of the event. The second subpanel shows the location of time–distance plots between slits P1 and P10. (b) Time–distance plots representing the POS transverse motion. (c) Image offsets related to co-alignment of the images where the spicule-type even is observed.

Spectral analysis is made on the transverse displacement of the structures. In this regard, it is worth noting that this analysis (see, for instance, wavelet diagrams) reveals a coherent motion of the spicules that cannot be reproduced by a random process like the one characterizing residual image aberrations induced by atmospheric turbulence and not corrected by the adaptive optics (AO) system. Further, the characteristic timescale of the seeing aberrations is of the order of 1–100 ms (see, e.g., Stangalini et al. 2017), depending on the wavelength, and thus is not consistent with the detected oscillations, which have much longer periods.

## 4. Case Studies

We focus on spicule-type events observed at ±1032, ±774, and ±516 mÅ in the Hα spectral line. Details of 30 events are given in Appendix A, and the time–distance diagrams representing POS motion are shown in Appendix B. These events are selected to show swaying motion that is (a) nonperiodic POS transverse motion, (b) periodic POS transverse motion in one of the line wings, (c) periodic LOS transverse motion, (d) nonperiodic POS and LOS transverse motions, or (e) periodic oscillations in both POS and LOS transverse motions.

We characterize the spicule-type events based on the criterion of whether they show periodic or nonperiodic signatures in the time–distance diagrams representing the POS displacement and/or in the LOS Doppler velocity. This POS motion is observed as a back-and-forth swaying motion performed by the spicule-type events. The POS periodic or nonperiodic motion is observed by an "S" shape in its time–distance diagram. Due to the nature of the spicule-type events, this POS transverse motion might not be oscillating with the same period or amplitude. However, in many cases we observe a "W" shape or "M" shape, which can suggest that the spicule-type event is undergoing more than one oscillatory cycle, but might not complete the second oscillation. In the LOS, this is detected as at least one sign inversion in the Doppler velocity images. We observe that in most cases the perturbations are nonperiodic where the oscillations last only one period. However, we identify multiple events that show periodic LOS and POS motions. We will illustrate the various combinations of POS and LOS signatures through several case studies.

*(a) Nonperiodic POS transverse motion.* Figure 5 shows spicule-type event #20 from Table A3 and shown in Figure B3. It belongs to data set 2 and is observed between 07:46:50 UT and 07:47:45 UT at −774 and −1032 mÅ in Hα. It is one of the longer-lasting events, with a lifetime and period of ∼80 s. We observe one cycle of oscillation that is apparent as an "S" shape in the time–distance diagrams. This spicule-type event is representative of spicules #2, #3, #4, #14 from Table A1 and #19 and #23 from Table A3.

*(b) Periodic POS transverse motion in either blue or red wings.* Figure 6 shows the evolution of spicule-type event #28 from Table A3 and Figure B4 observed between 07:36:06 UT and 07:37:02 UT. It belongs to data set 2. This event shows nearly two cycles of oscillation in the POS time series, with a time period of ∼20 s and amplitude of ∼100 km before disappearing. However, the oscillation shows considerable variation with time in amplitude and period. Figure 6 shows no distortions in the sequence of images, simultaneously verifying the co-alignment between the images. We are therefore confident in the true nature of this oscillation. This spicule-type event is representative of spicules #5, #6, #7, and #12 from Table A1 and spicules #3, #21, #24, and #25 from

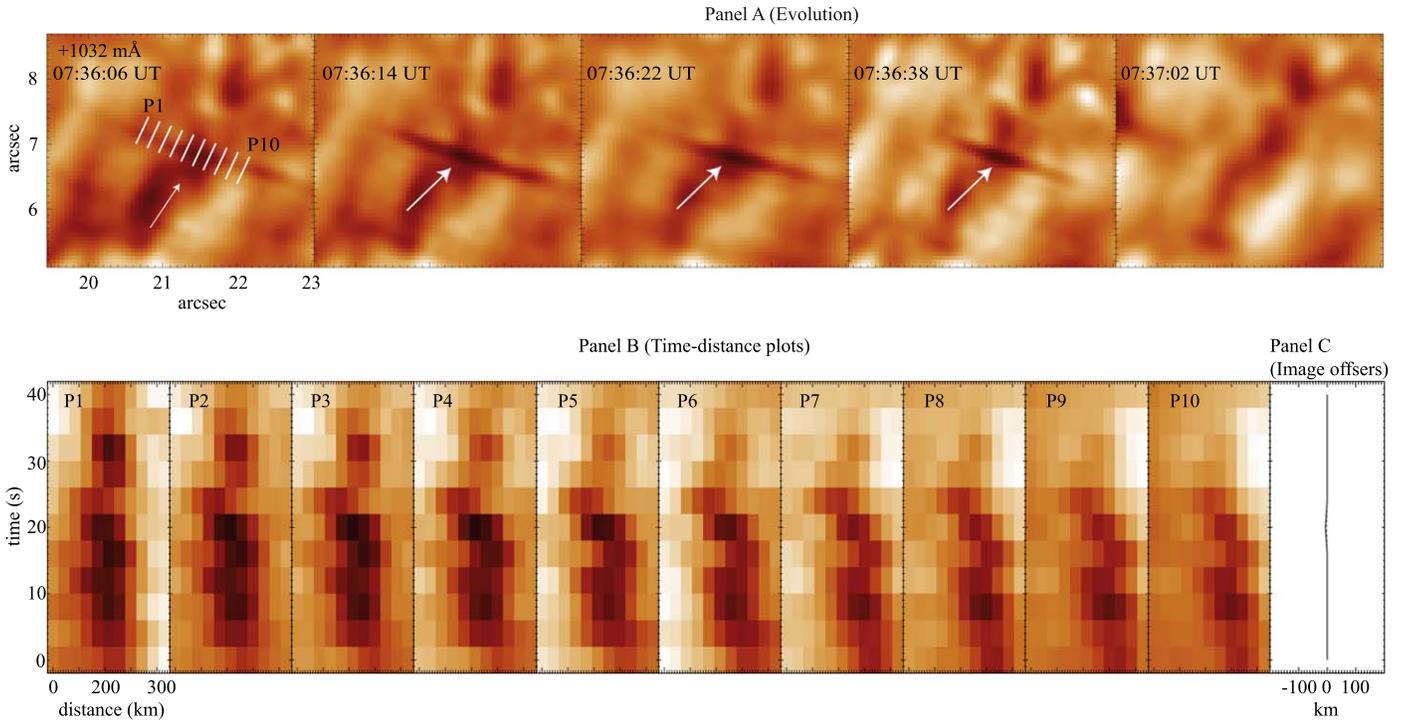

**Figure 6.** (a) Evolution of spicule-type event #28 from Table A3 and Figure B4, observed between 07:36:06 UT and 07:37:02 UT. White arrows indicate the location of the event. The second subpanel shows the location of time–distance slits P1–P10. (b) Time–distance plots along five slits. (c) Image residual jitter related to co-alignment of the images where the spicule-type even is observed.

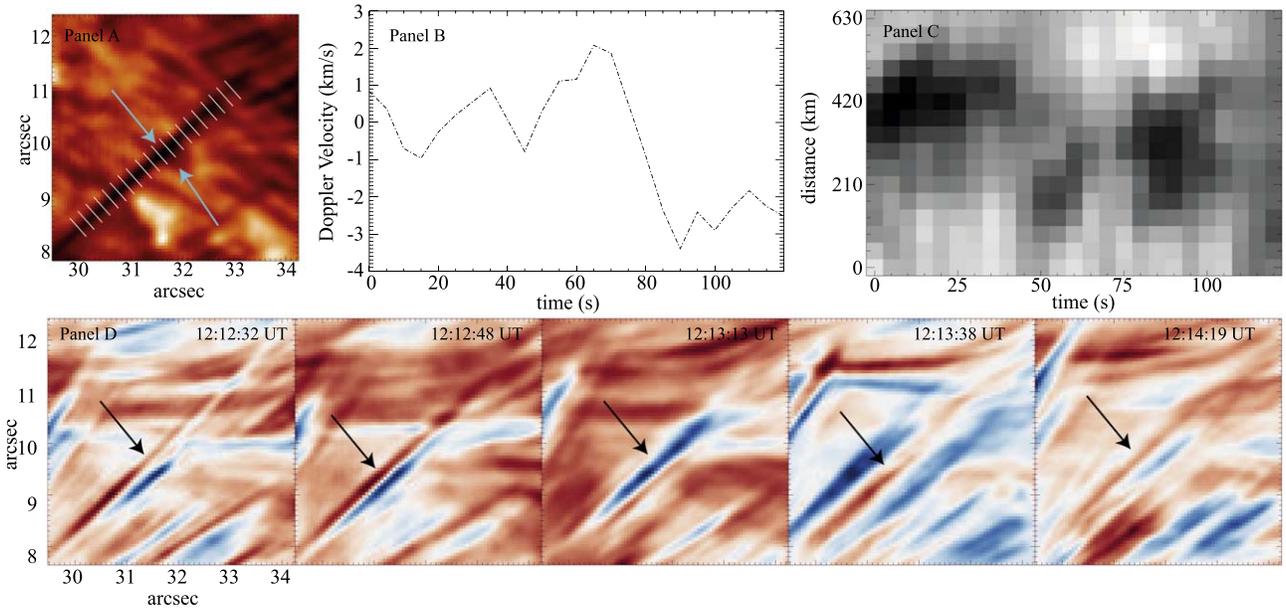

**Figure 7.** (a) Context image for spicule-type event #15 in Table A1, obtained at +774 mÅ observed between 12:12:32 and 12:14:19 UT on 2014 June 5. The arrows indicate the location of the spicule-type event. The location of slits long the length of the spicule-type event is shown as white solid lines. (b) Doppler velocity in km s$^{-1}$ at slit position 13. (c) Time–distance diagram corresponding to the POS motion. (d) Evolution of the Doppler velocities corresponding to the spicule. The red (blue) color corresponds to motion away from (toward) the observer.

Table A3. Note that some of these spicule-type events show 1.5 oscillations, depending on the lifetime of the event. However, we observe at least one complete cycle.

*(c) Periodic LOS transverse motion.* Figure 7 shows the evolution of spicule-type event #17 from Table A1. It is part of data set 1. This spicule-type event is observed in both red- and blue-wing positions of H$\alpha$ at ±774 and ±1032 mÅ. The spicule-type event is observed from 12:12:19 UT to 12:14:20 UT, i.e., for a duration of 120 s. This event starts appearing in the blue wing with appearances in the red wing at 12:12:32 UT (see Figure 7(d)), with alternative red- and blue-wing appearances thereafter. The cospatial and cotemporal appearance of events in the red and blue wings suggests that it is the same event but observed differently in two opposite line positions. We believe that such observations show LOS transverse motion and are observed as red–blueshifts in the Doppler velocity. The change in Doppler velocity computed at slit position 13 along the event, using moment analysis, is shown in Figure 7(b). The time–distance diagram of the corresponding POS signal does not show any definite motion (see Figure 7(c)).

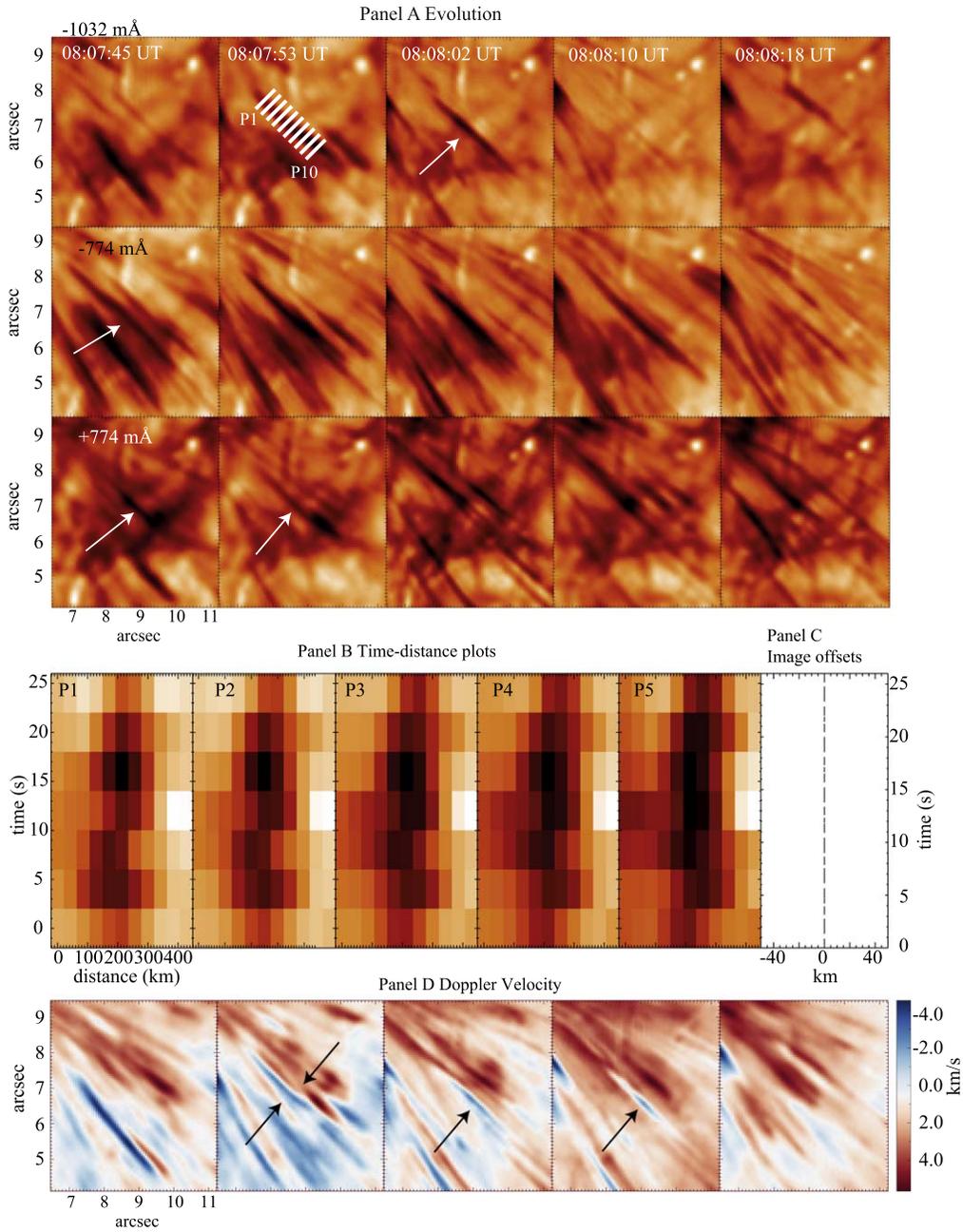

**Figure 8.** (a) Evolution of a spicule-type event (#16) in Table A3, observed between 08:07:45 UT and 08:08:18 UT on 2014 June 7. This spicule-type event is observed in the −1032 mÅ line position and at ±774 mÅ. White arrows show the location of the event, and the solid lines show the position of the slits. (b) Time–distance plots in the image-plane motion obtained at −774 mÅ. (c) Image offsets related to co-alignment of the images where the spicule-type event is observed. (d) Doppler velocity evolution.

*(d) Nonperiodic POS and LOS transverse motions.* Figure 8 shows spicule-type event #16 from Table A3. It belongs to data set 2 and is observed at ±774 and −1032 mÅ in H$\alpha$. The event lasts between 08:07:45 UT and 08:08:48 UT. This event shows a nonperiodic transverse oscillation in the image-plane direction as shown by the "S" shape in the time–distance diagram in the blue-wing positions −774 and −1032 mÅ. The LOS transverse motions show one sign change from blue to red to blue (see Figure 8(b)).

Figure 9 shows XWT plots for this. We can trace the POS motion in the blue-wing position −774 mÅ from the H$\alpha$ line center. We are able to observe one oscillatory period in both components of the transverse motion. Panel (a) shows the Doppler velocity, while panel (b) shows the image-plane displacement. The wavelet plots shown in the bottom panel of Figure 8 show a nominal period of around 20 s. The phase angle is shown by the phase vectors pointing toward the right. The XWT plots reveal in-phase behavior between LOS velocity and POS displacement. This spicule-type event is a representative of events #8, #11, #12, and #22 from Table A3.

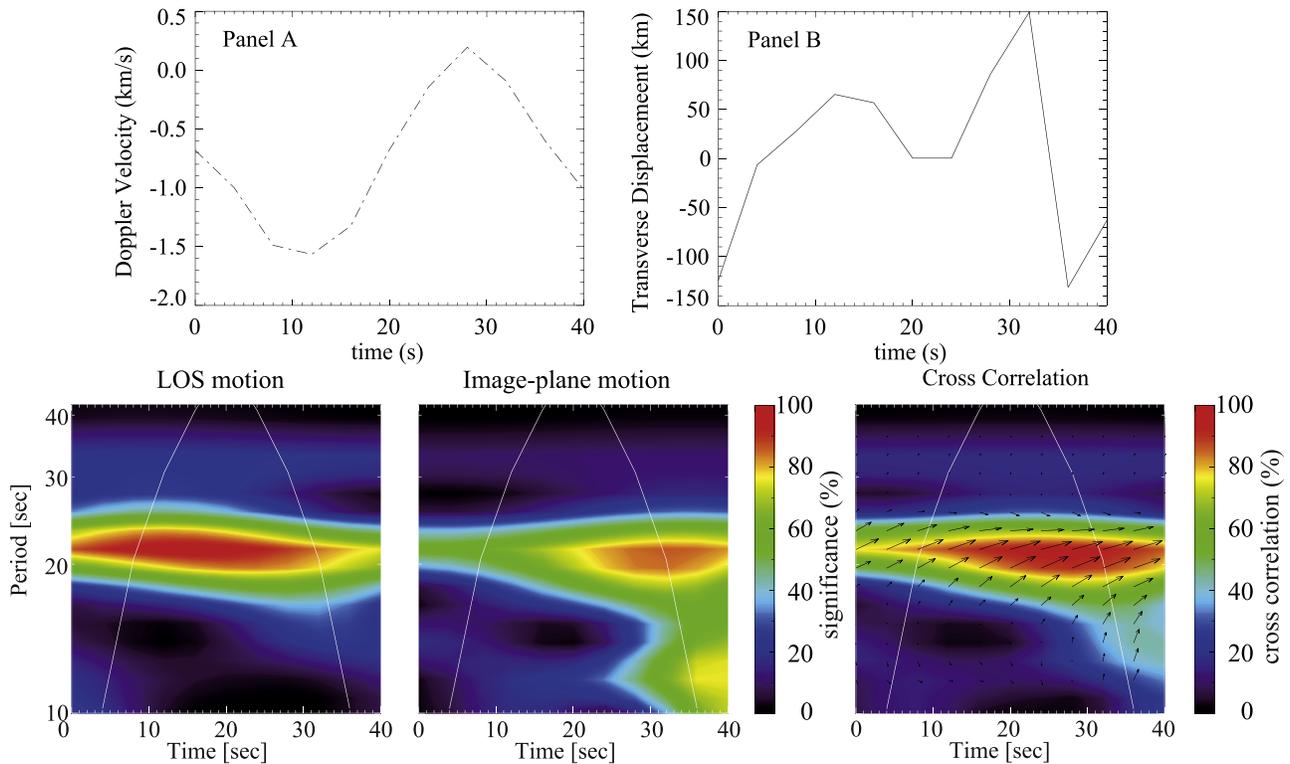

**Figure 9.** Phase relation corresponding to the spicule-type event #16 shown in Figure 8. (a) Doppler velocity variation showing the LOS motion. (b) Image-plane transverse displacement. Bottom row: wavelets obtained at the center of the spicule-type event. Left: Doppler velocity wavelet representing LOS motion. Middle: transverse displacement wavelet for the POS motion. Right: cross-correlation wavelet between two transverse motions. Red color represents regions with maximum significance, and black represents regions with minimum significance. Black arrows indicate through their length the correlation strength and through their direction the correlation phase.

*(e) Periodic oscillations in both POS and LOS transverse motions*. Figure 10 shows spicule-type event #20 from Table A3, as an example showing both POS and LOS oscillatory signatures (see Figure 10(a)). The event is observed between 08:03:09 UT and 08:04:45 UT at ±1032 mÅ from the Hα line center. The event is seen to oscillate with a period of approximately 70 s. However, the oscillation is only apparent in the blue-wing position, where the event lasts sufficiently longer (Figure 10). The LOS motion is observed as Doppler velocity variation, which shows a blue-wing feature with red-wing features observed at a distance of 1 pixel away, i.e., ≈50 km.

Figure 11 shows the phase relationship for the spicule-type event #20. We can trace the transverse direction in image plane corresponding to POS motion, in the blue-wing position −1032 mÅ from the Hα line center. We are able to observe one oscillatory period in both components of the transverse motion. Panel (a) shows the Doppler velocity and panel (b) the image-plane displacement. The period of this oscillation lies between 40 and 60 s. The oscillatory period for the LOS motion is about 50 s, whereas the POS motion shows one oscillation of about 40 s. Though the periods are offset, they do match at around 50 s when the two signals coexist. The XWT reveals in-phase behavior between the two time series.

Figure 12 shows spicule-type event #9 from Table A1, as an example of multiple closely aligned spicules that exhibit coherent transverse displacements. They are spatially separated by approximately 45 km. Panel (a) of Figure 12 shows the evolution of a spicule-type event in the red wing, between 12:21:59 UT and 12:24:06 UT at +774 mÅ and +1032 mÅ, and the event starts at 12:23:00 UT in the blue wing at −774 mÅ and −1032 mÅ. Multiple slits are drawn across the length of the spicule-type event to observe the POS motion (see slit positions in the second image of Figure 12(a)). The cospatial and cotemporal nature of the event suggests that the signatures in opposite wing positions correspond to the same event. The time–distance plots in Figure 12 show that there may be more than one cycle of oscillation in the blue wing at −774 mÅ with a period of ≈40 s with an amplitude of 100 km. The oscillation amplitude varies with time and is reduced to <100 km in the last 20 s of evolution. In the red wing we see a long oscillation with a period of about 80 s at the footpoint (or lower) end of the spicule with an amplitude of about 100 km. At the top end (or free end) of the spicule, the oscillation has a period of about 60 s with a much higher amplitude of 250 km. The occurrence of different periodicities between both wings indicates the presence of multiple spicule-type events aligned along the same LOS, which might follow a larger coherent structure.

In order to understand the LOS transverse motion, we show the Doppler velocity behavior computed using the moment analysis in panel (d). The spicule-type event appears redshifted at 12:22:10 UT and changes into a blueshifted feature at 12:23:05 UT. It then appears to show both a blue- and redshifted component of Doppler velocity until 12:23:20 UT.

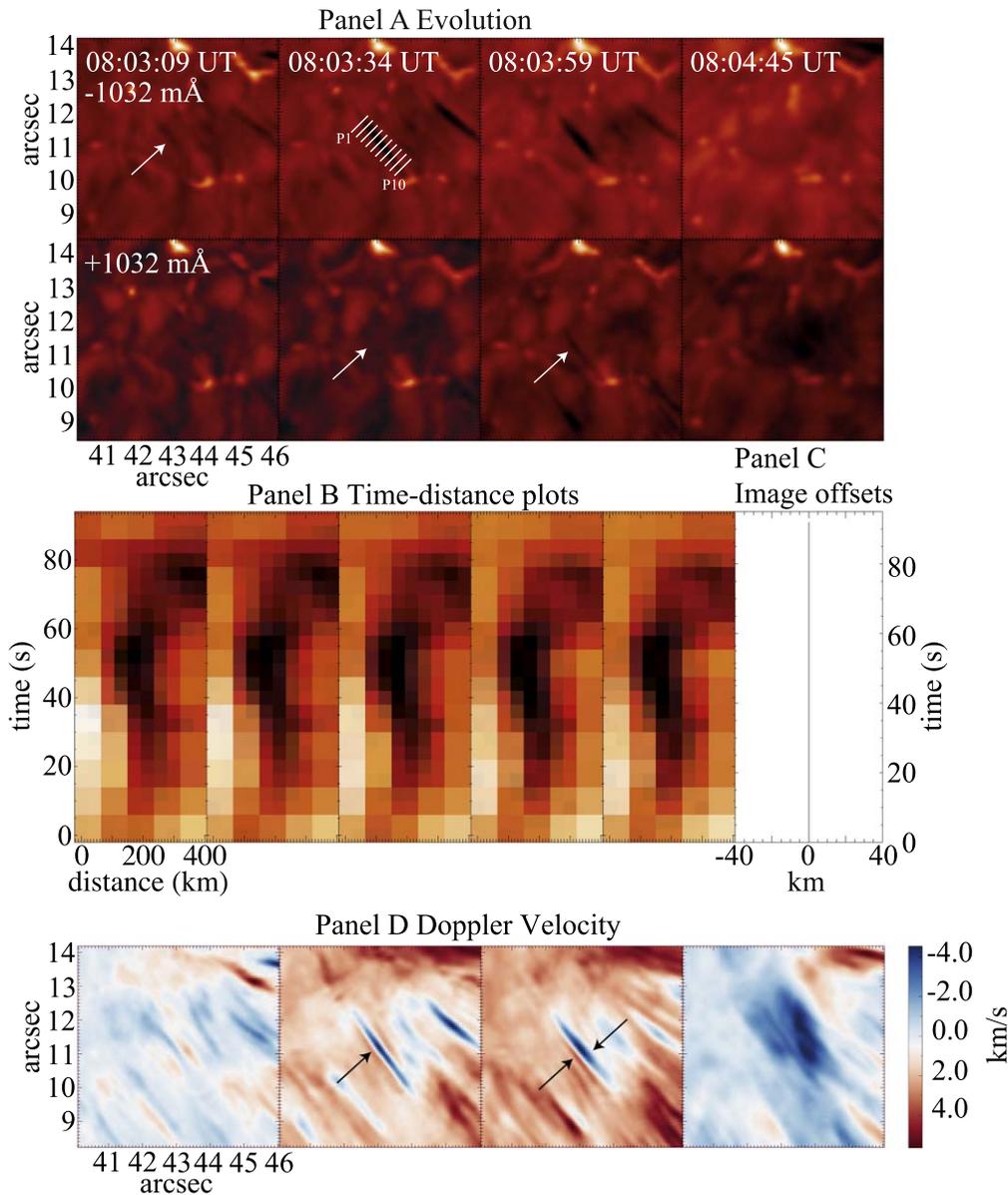

**Figure 10.** (a) Evolution of spicule-type event #20 from Table A3 between 08:03:09 UT and 08:04:45 UT in −1032 mÅ in Hα. White arrows show the location of the event. The second subpanel shows the location of time–distance slits P1–P10. The rightmost panel in panel (b) shows the seeing conditions during the event. (c) Image offsets related to co-alignment of the sequence of images where the spicule-type event is observed. (d) Evolution of the Doppler velocity corresponding to the spicule-type event.

The spicule-type event shows a redshift at the end of the evolution. The redshifted component of the spicule-type event lasts for ≈120 s, as compared to the blue-wing component that appears for ≈60 s. We notice that at 12:23:20 UT the spicule-type event shows relatively high Doppler width ∼36 km s$^{-1}$, indicating broadening of the line profile.

Figure 13 shows phase relationship plots for two spicule-type events. Both these spicule-type events show POS motion. The red-wing event shows a period of about 80 s in the LOS and image-plane motion as observed from the Doppler velocity and image-plane displacement plots shown in panels (a) and (b) of Figure 13. The XWT shows that the LOS velocity and POS displacement are approximately in phase. The blue-wing event is not observed as long as the red-wing event and starts after ≈50 s and at 1 pixel away from the red-wing event. The blue wing of the LOS velocity has a periodicity of around 30 s; the period of the POS displacement is a few seconds shorter. The XWT plot around that period shows that the LOS velocity and POS displacement are approximately in phase.

Figure 14 shows spicule-type event #8 from Table A1 and data set 1 as a second example of multiple closely aligned spicules that exhibit coherent transverse displacements. Panel (a) of Figure 14 shows the evolution of a spicule-type event at ±1032 mÅ starting at 12:00:44 UT and lasting until 12:03:10 UT. The event appears in the red wing at 12:00:54 UT and in the blue wing at 12:01:09 UT. The event then alternates between blue and red wings until 12:03:10 UT. During this evolution, the event shows a transverse oscillation in both the POS and LOS directions. The time–distance diagrams in Figure 14 show the presence of an oscillation in the POS with a period of approximately 100 s in both wings. The period of oscillation in LOS Doppler velocity is approximately 80 s.

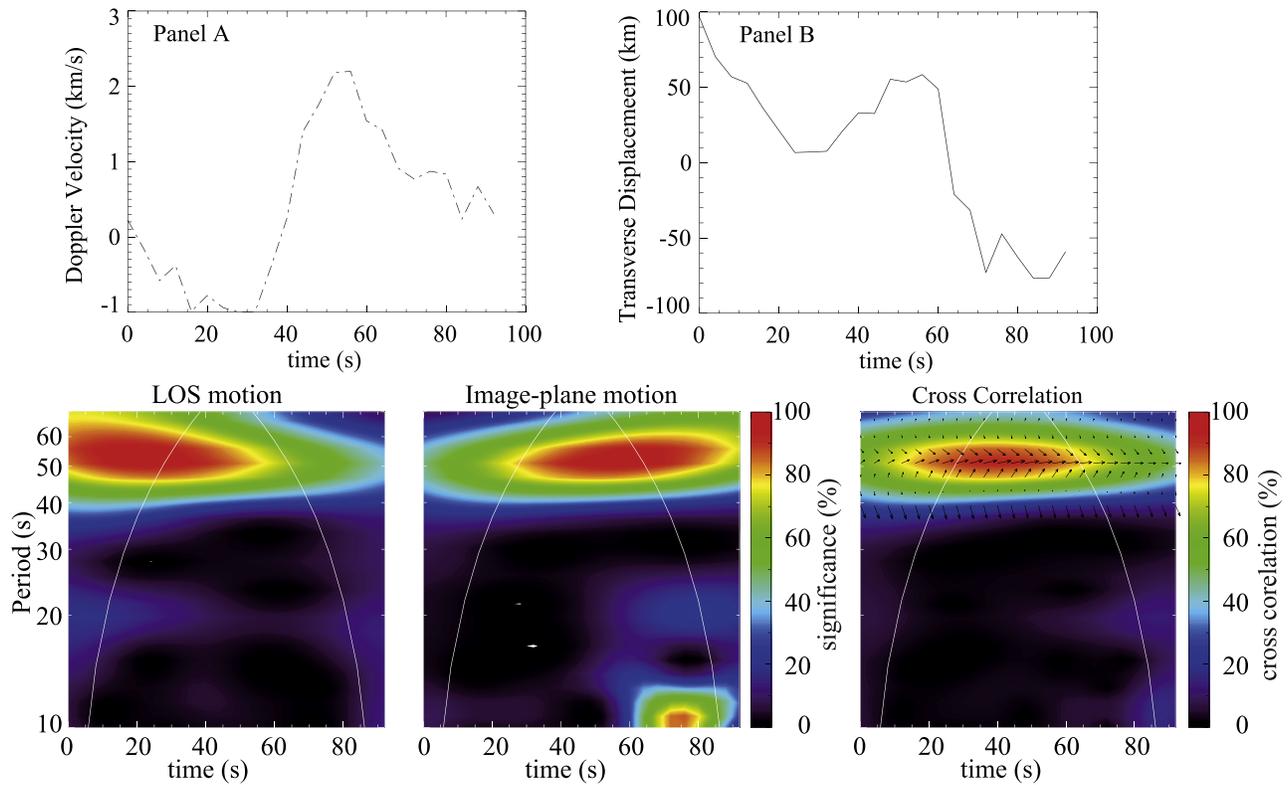

**Figure 11.** Same as Figure 9, but for spicule-type event #20.

Figure 15 shows the phase relationship plots for the two spicule-type events that are part of event #8, cospatial in the blue–red wing but not cotemporal as shown in Figure 14. These spicule-type events are best observed at ±774 and −1032 mÅ from the Hα line center. In the red-wing position, the event shows an oscillatory behavior in the image plane with a period between 60 and 80 s; the top two rows of Figure 15 show this behavior. The LOS motion has a period of 100 s. There are additional smaller oscillations shown in the time–distance plot of Figure 14, which are enhanced in the POS wavelet plots (see the second wavelet plot, which shows the transverse displacement). The XWT shows an in-phase correlation at the period of ≈60 s between POS displacement and LOS velocity. In the blue-wing position, the event shows one nonperiodic in-phase oscillation of ∼70 s in both POS and LOS transverse directions.

### 5. Statistics of Spicule Oscillations

The distributions of the observed physical properties of the events such as length and lifetime are shown in panels (a) and (b) of Figure 16. The length is the maximum length of the observed spicule during its evolution at the line position of the Hα line profile where the lifetime is computed. The lifetime is defined as the maximum duration that the event is observed in all of the Hα wavelength positions. The remaining analyses of the spicule-type events are performed at the line position where the feature is most prominent. From the observations, the lifetime of these spicule-type events is on average 60 s and the length is 3000 km with width of 100 km. We focus on events that show at least one period of oscillation in the POS direction at any line position along the Hα line scan.

The distributions of the properties of the POS transverse oscillations, such as amplitude, period, and velocity, are shown in Figures 16(c)–(e). The average displacement amplitude of the image-plane transverse motion is 100 km. The periods of oscillations are between 16 and 100 s and are shorter than the chromospheric cavity 3-minute acoustic period. Equivalently, the frequencies lie between 10 and 63 mHz, about an order of magnitude higher than the chromospheric natural frequency of 5 mHz (e.g., Fleck & Schmitz 1991; Hansteen 1997). As such, we designate these as high-frequency oscillations. The transverse velocity ranges between 5 and 40 km s$^{-1}$, with an average around 20 km s$^{-1}$.

In panel (f) we explore the relation between lifetime and period. Since for most cases the lifetime is only marginally longer than the period, we only see one or a few cycles of oscillation in each event. There is a lower cutoff in that there are no events with an oscillation period less than its lifetime. This is a selection effect because we chose to study only events with an oscillatory displacement pattern.

In panel (g) we plot the oscillation period versus the length. There is no clear correlation. If we assume that all events have similar magnetic field strengths and densities, then this would suggest that spicule length is not related to the oscillation wavelength. Therefore, we expect the oscillation to be supported on a scale larger than the spicule-type event itself, thus hinting toward the presence of multiple coherent structures within a larger structure and acted upon by an external trigger.

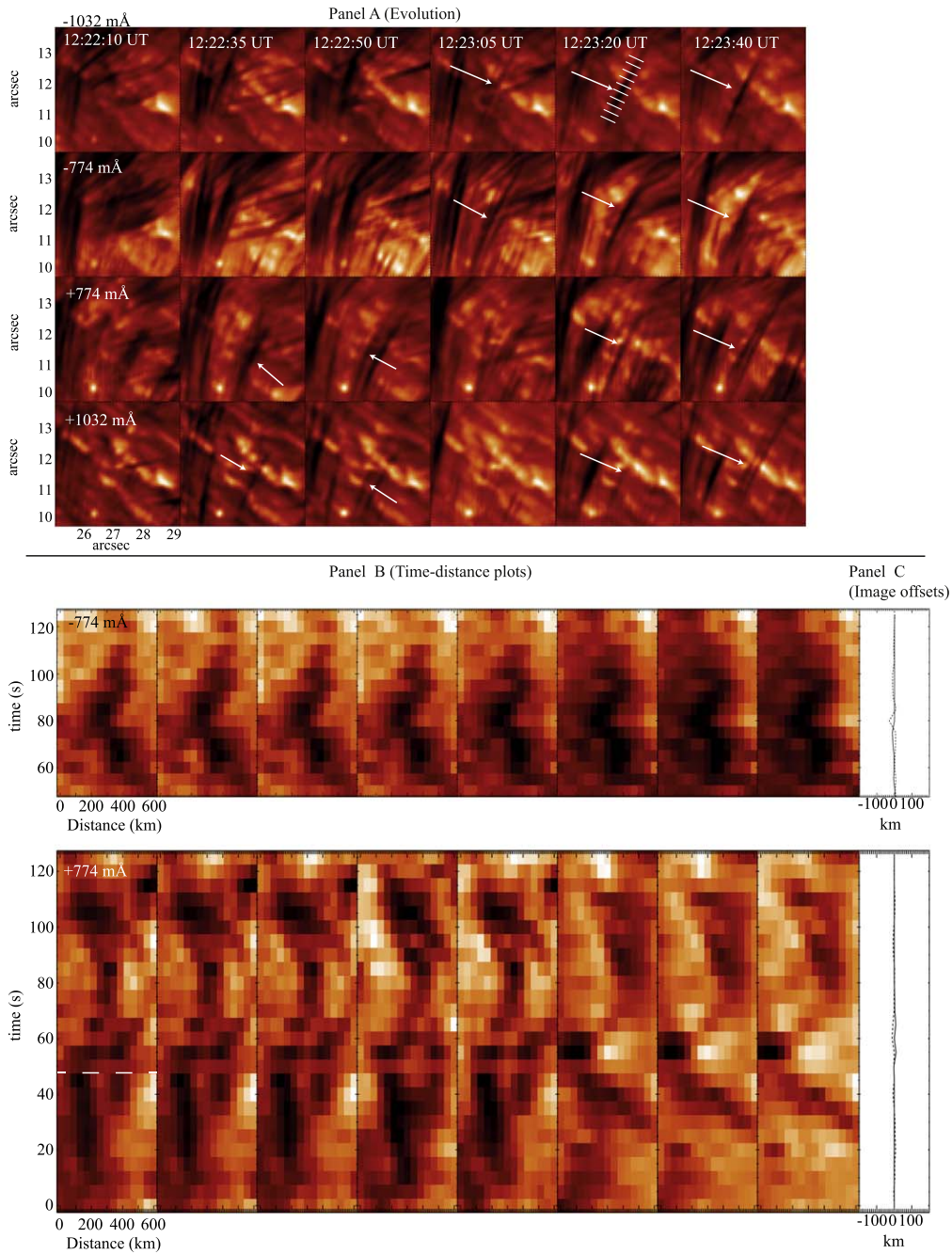

**Figure 12.** (a) Evolution of spicule-type event #9 between 12:22:10 UT and 12:23:40 UT at ±1032 and ±774 mÅ. White arrows indicate the location of the spicule. The second-to-last subpanel of the top row shows the location of time–distance slits P1–P10. (b) Time–distance plots showing image-plane motion in two wing positions at ±774 mÅ. The white dashed line on the time–distance plot of the red-wing position +774 mÅ shows the start of the oscillation in the blue-wing position at −774 mÅ. (c) Image offsets related to co-alignment of the sequence of images where the spicule-type event is observed.

Lastly, we plot in panel (h) the relationship between the transverse displacement, oscillation amplitude, and the period of the oscillation. We find no particular correlation, which indicates that the excitation mechanism responsible for the oscillations is homochromatic in this period range. Furthermore, the displacement amplitudes are an order of magnitude smaller than the spicule length, which in turn are much smaller than the oscillation wavelength. In this respect, the oscillations can be considered to be linear. However, the displacement amplitudes are of the same order as the width of the spicules and may in that respect be considered nonlinear only if the oscillation is considered restricted to the structure itself.

### 6. Discussion

In JS16, the focus was on the sudden appearance and disappearance of a magnetic-structure spicule entering our FOV and scanning through a narrowband filter of CRISP. These spicule-type events were observed at ±516, ±774, and ±1032 mÅ, respectively. Here we focus on 30 spicule-type events with average observed lengths of about 3000 km, width (or diameter) of 100 km, and lifetimes of about 60 s. We report all spicule-type events with high-frequency oscillations with an average frequency of between 10 and 63 mHz for the two data sets, respectively. The mean velocity amplitude of the transverse oscillation in the POS is around 18 km s$^{-1}$, with a few events higher than 60 km s$^{-1}$. This is comparable with both the chromospheric acoustics speed and the propagation speed of the fast magnetoacoustic wave fronts. Therefore, these oscillations may be regarded as nonlinear in nature. In several cases, we find events that are spatially close but visible in opposite wing positions. They have different evolution. We also note a small temporal offset in occurrence in the red-wing and blue-wing positions at ±516, ±774, and ±1032 mÅ, respectively.

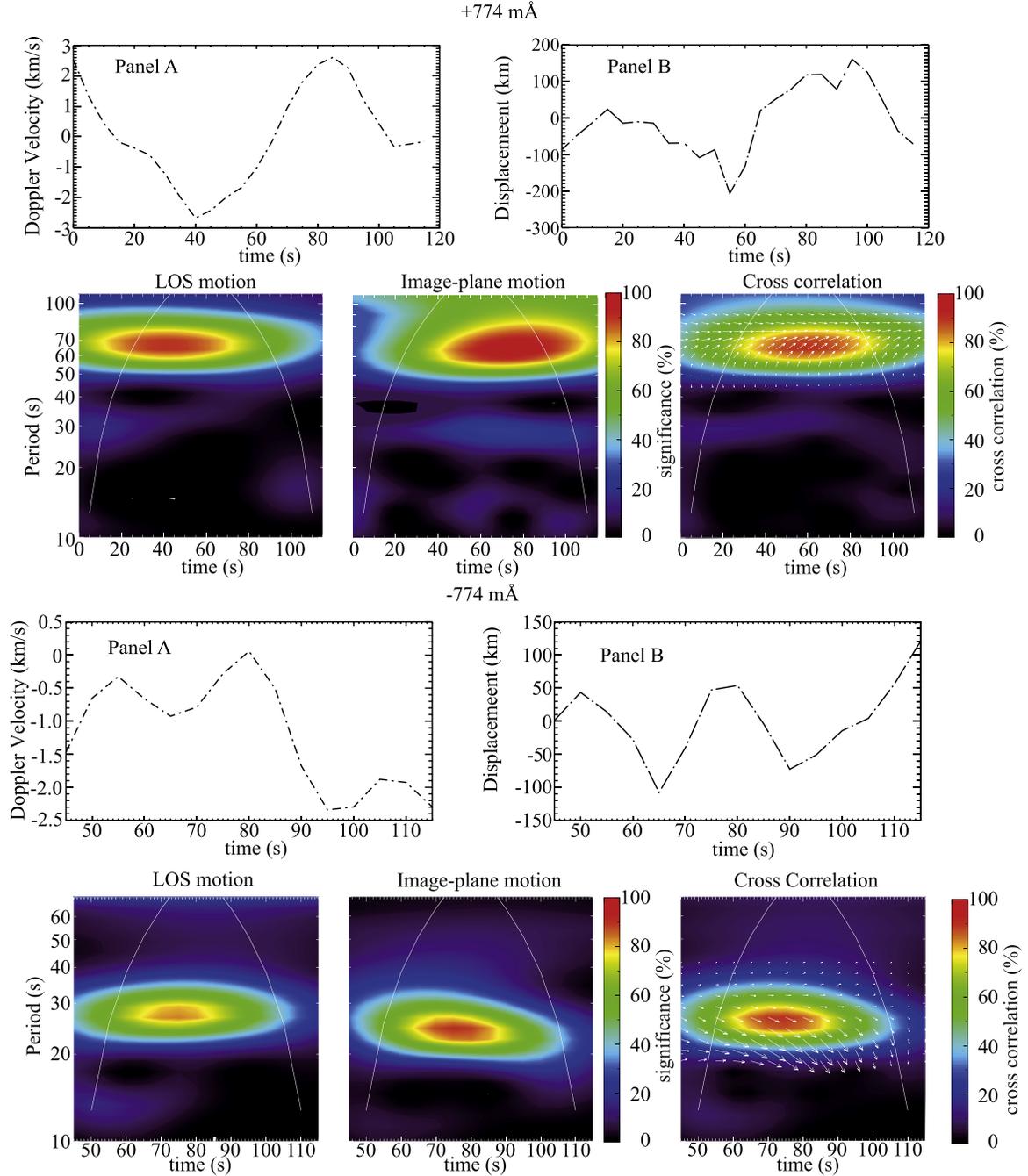

**Figure 13.** Phase relationship plots using wavelets obtained at the middle position along the two spatially close spicule-type events #9 shown in Figure 12. The top two rows show time series of the event from the center of the spicule-type event visible in the red wing at +774 mÅ. (a) Doppler velocity variation showing LOS motion at the position of this event. (b) LOS transverse displacement of this event. Bottom row: three wavelet plots for the two time series. Left column: wavelet plot for the LOS Doppler velocity time series. Middle column: wavelet plot for the POS displacement time series. Right column: cross-correlation wavelet plot between the two time series. Black arrows indicate through their length the correlation strength and through their direction the correlation phase. The two bottom rows show the same plots but for the event visible in the blue-wing position at −774 mÅ.

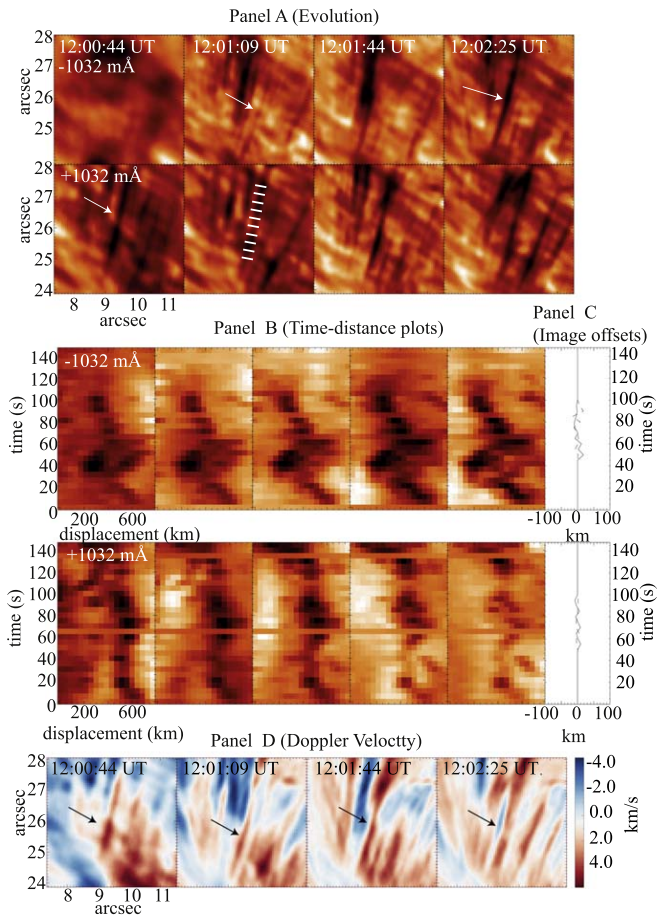

**Figure 14.** (a) Evolution of spicule-type event #8 between 12:00:44 UT and 12:03:10 UT at −1032 mÅ in Hα. White arrows indicate the location of the event. The second subpanel shows the location of time–distance slits. (b) Time–distance diagrams showing evolution at ±1032 mÅ. (c) Image offsets related to co-alignment of the sequence of images where the spicule-type event is observed. (d) Evolution of Doppler velocity.

The periods of oscillations are nearly of the same order of magnitude. However, we note that the limited lifetime of these spicule-type events hampers the detection of periodicities longer than about 100 s. The solid red line in Figure 16(h) suggests that there is a lower limit to the observed period of oscillation imposed by the observed lifetime. During the lifetime of a spicule, not many cycles of the oscillations are visible, and the mean value of 60 s cannot be regarded as a characteristic period. Furthermore, we have found no correlation between oscillation period and length. This indicates that the length is not related to the oscillation wavelength. The oscillation must be supported by a structure larger than the spicule-type event itself. Furthermore, we have not detected any oscillation phase difference between different locations along the spicule structure. This suggests that the wavelength parallel with the structure is much longer than the spicule length. The parallel phase speed will be larger than $L/P \approx 50$ km s$^{-1}$. Large uncertainties in amplitude and periods deduced from the POS transverse motion could originate from the closeness of spicule-type events to nearby events. To resolve this entanglement, observations at higher spatial resolution are needed, e.g., from the 4 m class Daniel K. Inouye Solar Telescope (DKIST; Rast et al. 2021).

We decompose the high-frequency motion into two components along the observational channels of the POS and LOS. We show that sinusoidal patterns in time–distance diagrams correspond to (periodic) transverse displacements in the POS. The LOS oscillations are reported as a change of sign in the Doppler shift. The spicule-type events show either a blue–red–blue sign change or a red–blue–red sign change (see Figures 5, 8, 12, 14, and 10). The relation between these two components is explored using phase relations between time series extracted at the center of the event. Cross-correlation wavelets show that the POS displacement and LOS velocity associated with an oscillation are in phase. We interpret this phase behavior as indicative of a helical oscillation that follows an elliptical polarization pattern as follows. Consider the POS and LOS displacements to be of the form $\xi_{POS} \sim \cos(\omega t + \phi_{POS})$ and $\xi_{LOS} \sim \cos(\omega t + \phi_{LOS})$. The wave mode is linearly polarized if the two displacement time series are either in phase or in antiphase, i.e., $\phi_{POS} = \phi_{LOS}$ or $\phi_{POS} = \phi_{LOS} + \pi$, respectively. The wave mode is circularly (or generally elliptically) polarized if there is a ±90° lag between both time series, i.e., $\phi_{POS} = \phi_{LOS} \pm \pi/2$. For the phase difference between the POS displacement and the LOS velocity, which is of the form $v_D \sim \sin(\omega t + \phi_{LOS})$, elliptical polarization translates into an in-phase or antiphase relation. Figure 17 traces the motions of the spicule-type events in terms of the POS and LOS velocities. A clear rotational pattern is visible superimposed on drift motions. The POS transverse speed is at least an order of magnitude larger than the LOS speed. However, the suppression of the LOS signal could be due to multiple factors such as opacity differences in two opposite wing positions and superposition of a static background in the LOS.

We conclude that the observed periodic transverse motions in short-lived spicule-like events arise as a result of a helical Alfvénic (kink) wave that is elliptically polarized. For a helical kink wave, the POS displacement and LOS velocity are expected to oscillate in phase. Zaqarashvili & Skhirtladze (2008) studied the theoretical aspects of the helical motion in a magnetic flux tube. They showed that if two or more kink waves with different polarization are excited in the same thin tube, then the superposition may set up helical motions of magnetic flux tubes in the photosphere/chromosphere. Helical waves were also observed in chromospheric small-scale magnetic concentrations (Sharma et al. 2017). However, in our examples each spicule-type event that is part of the spicule structure will have a helical kink wave. Figure 17 shows the projected motion of six spicule-type events. In all cases, the presence of helical motion is visible (indicated by black arrows) superimposed on top of a slower trend. This trend causes the helical motion to not resemble a circular or elliptical path centered on the origin, but as stretched and displaced. However, the local reversal in motion remains apparent. For the spicule-type event reported in Figure 12, we observe that the two spicules tracked indicate that the direction of helical motion can be in both the clockwise and anticlockwise direction (see panels (a.1) and (a.2) of Figure 17). For the spicule-type event reported in Figure 14, we observe that the helical motion is in the same clockwise direction (see panels (b.1) and (b.2) of Figure 17).

We also conclude that this helical oscillation is likely excited by fast magnetoacoustic wave packets that travel across the spicule structures. This is supported by the presence of wave fronts in time–distance plots for an image cut taken across multiple spicules. We have estimated the POS projected propagation speed to be in the range 13–18 km s$^{-1}$. It is also supported by the lack of evidence of a phase difference between the transverse motion measured at any point along the spicule. This suggests that the parallel wavelength is much longer than the structure. This particular piece of evidence alone does not allow us to distinguish between a standing wave and a wave whose component of propagation along the structure is too fast to be detected. However, the large range of reported parallel propagation speed is consistent with the latter scenario (Sekse et al. 2013b). The presence of standing waves in slower versions of the events is discussed by Sekse et al. (2013a). For a time cadence of 4–5 s, we require a parallel phase speed of the order of 450 km s$^{-1}$ to cover a typical spicule length of 3″ (2200 km) in that time. The long parallel wavelength and absence of phase difference along the spicule are also consistent with our favored scenario of the wave propagating nearly perpendicular to the spicule axis at speeds of 13–18 km s$^{-1}$. The wave front would cross each point along the spicule almost instantaneously, and the parallel (projected) wavelength will be much longer than the actual wavelength that will be of the order of a few hundred kilometers. This wave will be fast magnetoacoustic in nature. The velocity amplitudes of the spicules are comparable with the fast wave propagation speed, which suggests that the spicule oscillations are nonlinear. The passing waves may be the drivers of the spicule oscillation. We hypothesize that the compression and rarefaction of the passing magnetoacoustic wave may influence the appearance of spicule-type events, not only by contributing to moving them in and out the wing of the spectral line but also through the creation of density enhancements and an increase in absorption in the H$\alpha$ line.

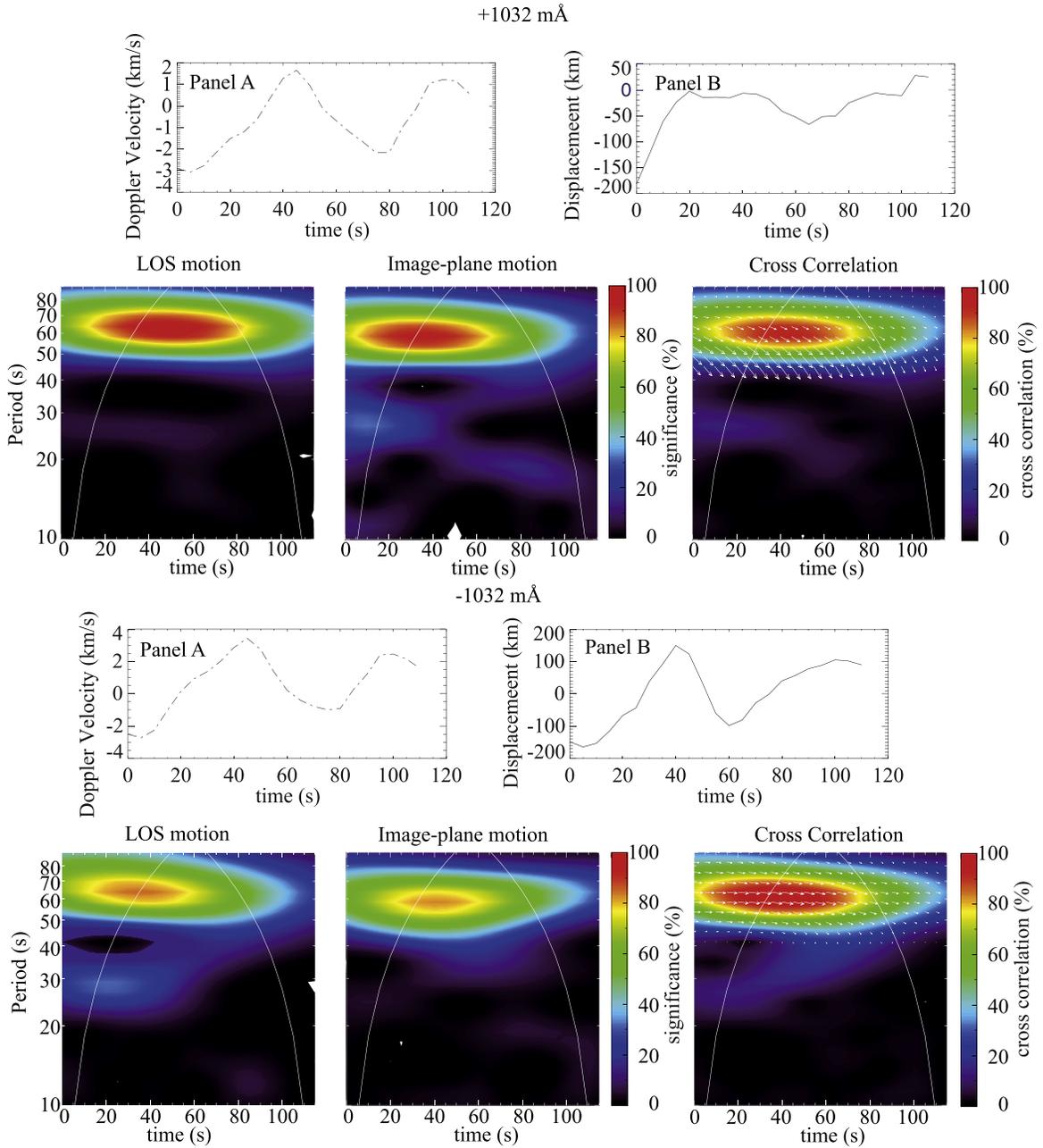

**Figure 15.** Same as Figure 13, but for event #8.

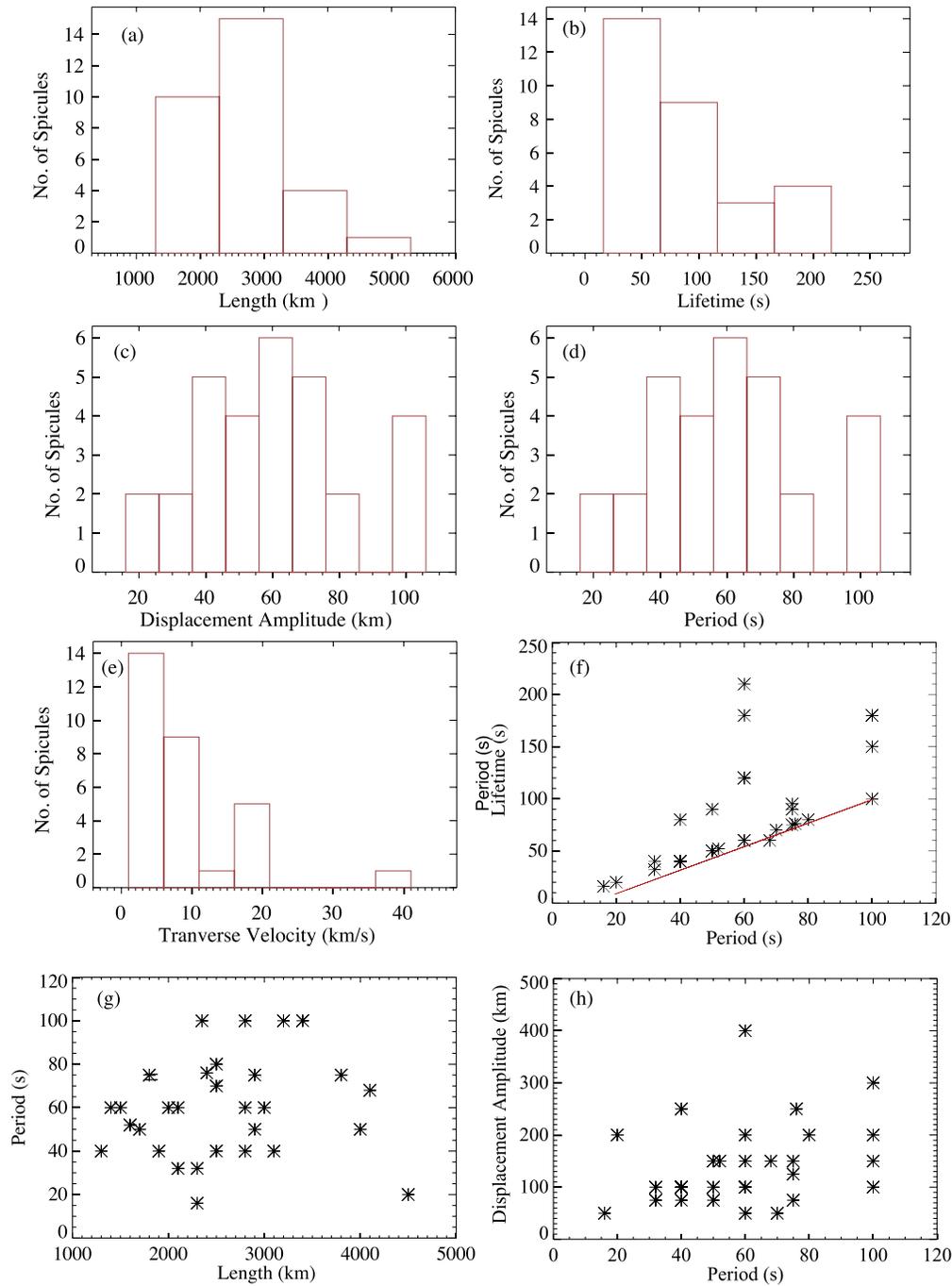

**Figure 16.** Statistics corresponding to 30 observed spicule-type events. (a) Histogram of the length. (b) Histogram showing the lifetime. Panels (c)–(e) show histograms representing the period, amplitude, and velocity of the POS transverse motion, respectively. Panels (f)–(h) show the relationship between quantities such as the period and lifetime, (POS transverse velocity)/length, and amplitude.

A similar density dominance was reported in H$\alpha$ fibrils by Leenaarts et al. (2012). It is unclear what physical mechanism can cause this at the timescales we see. Another possibility could be that the superposition of many different wave fronts of longer-period magnetoacoustic waves propagating in different directions leads to local oscillatory behavior. Another interpretation of the observational behavior (related to the red–blueshifts) could be associated with wave-induced Kelvin–

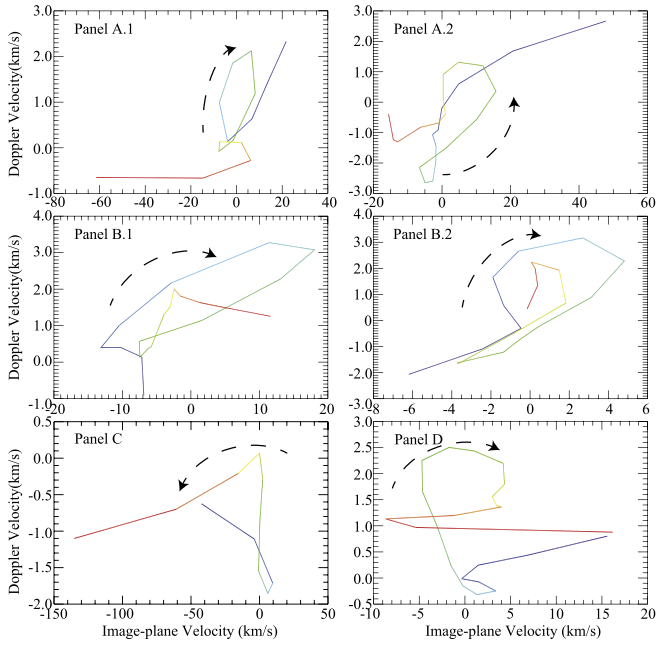

**Figure 17.** Projected motion of the position of spicules as a function of the POS and LOS transverse velocities. Time progression is indicated with color, from blue to red. Panels (a.1) and (a.2) correspond to spicule-type event #9 from Figure 12. Panels (b.1) and (b.2) correspond to spicule-type event #8 from Figure 14. Panel (c) shows this relationship event #1 from Figure 8, and panel (d) shows this relationship for spicule-type event #7 from Figure 10.

Helmholtz rolls, where the red–blue–redshifts observed produce a similar observational pattern observed in Doppler velocity plots (Antolin et al. 2018).

Putting all the observational evidence together, the oscillatory part of the motion of the spicule would follow a helical path. We have depicted this scenario in three dimensions in Figure 18. It shows one spicule at four stages of its helical motion. It shows that when the spicule is at the apex or bottom of its motion it exhibits no LOS motion and results in no net displacement or Doppler shift. There is a POS velocity, but that cannot be observed spectrally. It can only be deduced from the POS displacement. When the spicule has the largest POS displacement, to one or the other side, then there is also a Doppler shift. This illustrates visually how time series of displacement and Doppler shift are in-phase (or antiphase) for a helical motion.

Questions remain on how the signal from such high-frequency waves propagates across different solar atmospheric layers, suggesting the need for simultaneous observations in the transition region. Martínez-Sykora et al. (2015) suggest the presence of UV explosive events using the Interface Region Imaging Spectrograph (IRIS; De Pontieu et al. 2014b) as a possible link. In the near future, next-generation telescopes such as the 4 m class DKIST should be able to resolve these spicule-type events across multiple wavelengths and shed further insights over physical

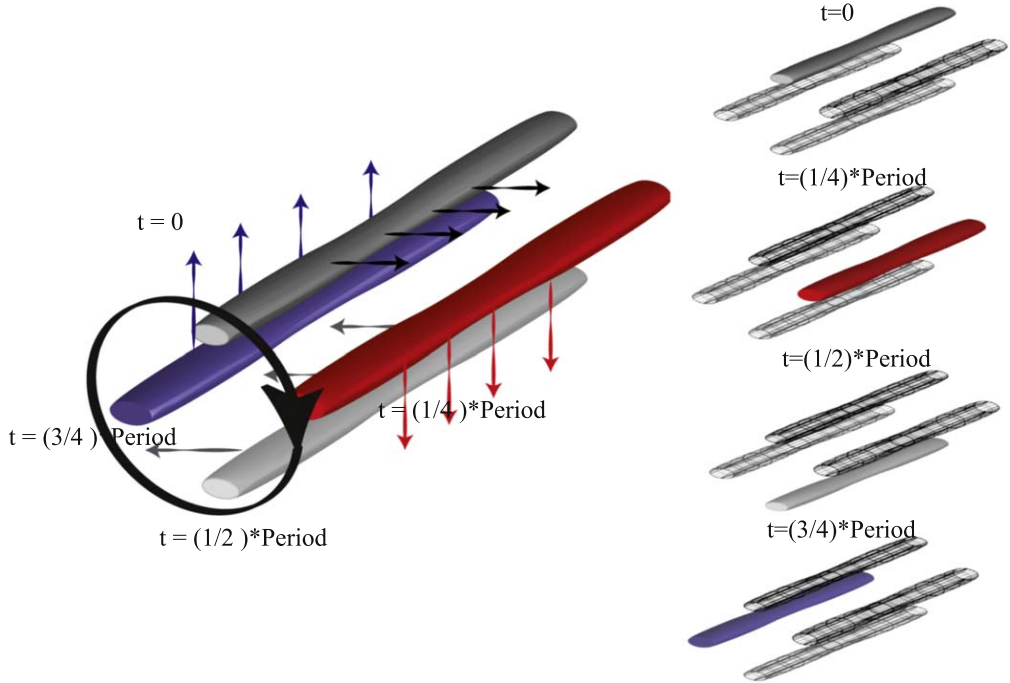

**Figure 18.** 3D projection of the scenario of a spicule undergoing a helical oscillation of period $P$, shown at four stages of its motion. At $t = 0$, the spicule is shown in gray as it exhibits no Doppler shift. However, its POS velocity is maximal to the right. At $t = 0.25P$ the spicule is shown in red, as there is a Doppler shift due to motion away from the observer and lower into the atmosphere. There is no POS velocity, and the spicule has reached its maximum POS displacement. At $t = 0.5P$, the spicule again depicted in gray as Doppler shift is absent, and there is a maximum POS velocity to the left. At $t = 0.75P$, the spicule is shown in blue, as there is a Doppler shift due to motion toward the observer and higher up in the atmosphere. The POS velocity is again zero, and the spicule has reached a maximum displacement.

processes that can cause such observational signatures associated with high-frequency motions.

The authors thank the anonymous referee for the invaluable comments and time. J.S. is funded by NSF grant No. 1936336. Armagh Observatory and Planetarium is grant-aided by the N. Ireland Department of Communities who part-funded the observing campaign. The Swedish 1 m Solar Telescope is operated on the island of La Palma by the Institute for Solar Physics of Stockholm University in the Spanish Observatorio del Roque de los Muchachos of the Instituto de Astrofísica de Canarias. J.S. would like to thank Dr. Eamon Scullion for assistance in data reduction. We would also like to thank STFC for PATT T&S and the Solarnet project, which is supported by the European Commissions FP7 Capacities Programme under Grant Agreement No. 312495 for T&S. J.G.D. would like to thank the Leverhulme Trust for an Emeritus Fellowship.

## Appendix A
## Summary of Events

We summarize the 30 spicule-type events analyzed for this study taken from two data sets, in the following tables (Tables A1–A4). We show a sample image when the spicule-type event is best observed. The lifetime (s) of the spicule-type event corresponds to the total time that the event was observed in the best observable wavelength. This is usually at $\pm 774$ mÅ from the H$\alpha$ line center. The period of swaying motion observed in the POS corresponds to one back-and-forth motion observed by an "S" shape. See Appendix B. This period is computed at $\pm 774$ mÅ. Comments are based on the location along the H$\alpha$ line profile.

Table A1
Table Showing Spicules Selected from Data Set 1

| Case no | Time (hh:mm:ss UT) | H$\alpha$ image | Comment |
|---|---|---|---|
| 1 | 11:57:31 | | Blue wing spicule-type event observed at -1032, -774 and -516 mÅ from the H$\alpha$ line centre. Seeing deteriorates during the evolution. Lifetime of the event is 210 s. |
| 2 | 12:17:21 | | Blue wing spicule-type event observed at -1032, -774 and -516 mÅ from the H$\alpha$ line centre. Lifetime is 60 s and observed swaying motion lasts for 60 s. |
| 3 | 12:19:53 | | Spicule-type event present in blue wing positions -516, -774 and -1032 mÅ from the line centre. Spicule-type event lasts longer at -516 mÅ. Lifetime is 95 s and observed swaying motion last for 95 s. |
| 4 | 12:16:46 | | Spicule-type event present in blue wing positions -516, -774 and -1032 mÅ from the line centre. Event lasts longer at -516 mÅ. Lifetime is 70 s and observed swaying motion last for 70 s. However there is recurring activity observed in this location, suggesting that the event could be recurring. |
| 5 | 12:19:23 | | Spicule-type event present in blue wing positions -516, -774 and -1032 mÅ from the line centre. Event lasts longer at -516 mÅ. Lifetime is 180 s. The oscillation present is irregular and show different periodicities of 70 s and 100 s. Observed at the same location as spicule-type event 4. |
| 6 | 12:18:02 | | Spicule-type event present in blue wing positions -516,-774 and -1032 mÅ from the line centre. Lifetime is 180 s. The oscillation present is irregular and show different periodicities of 50 s and 100 s, similar to spicule-type event 5. It is also observed at the same location as spicule-type events 4 and 5. |
| 7 | 12:17:21 | | Spicule-type event present in blue wing. positions -516, -774 and -1032 mÅ from the line centre. The event last 180 s and shows irregular oscillations. |
| 8 | 12:00:54 | | Spicule-type event is present in both wings at ± 516, ± 774 and ± 1032 mÅ from the line centre. It lasts for 100 s with swaying motion completing one back and forth motion in 60 s. |
| 9 | 12:23:00 | | Spicule-type event is mainly observed in red wing positions, +516, + 774 and + 1032 mÅ from the line centre. It lasts for 100. |
| 10 | 12:22:45 | | Blue wing spicule-type event observed at -516, -774 and -1032 mÅ. Lifetime is 40 s and observed swaying motion last for 40 s. |
| 11 | 12:03:00 | | Spicule-type event present in both red and blue wings at ± 516, ± 774 and ± 1032 mÅ from the line centre. Spatial offset in both the wings. Red wing signature last longer. Lifetime of the spicule-type event is 90 s and we see a 50 s oscillation at the start of the feature. |

**Note.** Column (1) shows a unique number corresponding to the events. Column (2) shows the time corresponding to the first appearance of the event. Column (3) shows the H$\alpha$ image. All images shown in Column (3) are obtained at +774 mÅ or −774 mÅ. Column (4) describes the events in detail.

Table A2
Table Showing Spicules Selected from 2014 June 05 Data Set

| Case no | Time (hh:mm:ss UT) | Hα image | Comment |
|---|---|---|---|
| 12 | 12:02:35 | 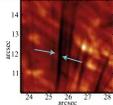 | Spicule-type event present in both red and blue wings at ± 516, ± 774 and ± 1032 mÅ from the line centre. However, signatures in both the wings have a spatial offset. Furthermore, the blue wing signature appears for fewer time-steps than the red wing signature. |
| 13 | 12:22:55 | 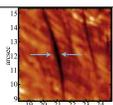 | Spicule-type event present in the red wing at +516, + 774 and + 1032 mÅ from the line centre. Lifetime is 125 s and period is 60 s. This spicule-type event shows peculiar oscillation, one end of the spicule shows larger amplitude of oscillation than the other. |
| 14 | 11:56:05 | 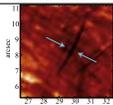 | Red wing event at observed at +516, + 774 and + 1032 mÅ from the line centre. Lifetime is 30 s and with swaying motion completing one back and forth motion in 30 s. |
| 15 | 12:12:17 | 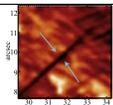 | This spicule-type event's motion shows non–periodic motion LOS. It is described in Fig 7. |

Table A3
Table Showing Event Selected from Data Set 2

| Case no | Time (hh:mm:ss UT) | Hα image | Comment |
|---|---|---|---|
| 16 | 08:07:41 | 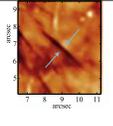 | Red and Blue wing event observed at ± 774 mÅ and 1032 mÅ, with lifetime of the event and LOS period of 25s. Red wing signature weaker in intensity than the blue wing signature. However the shift in LOS allows for phase relationship discussed in Fig 8. |
| 17 | 07:36:56 | 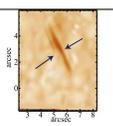 | Spicule-type event present in both red and blue wings at ± 774, and ± 1032 mÅ, from the line center. Swaying motion is only observed in the blue wing. Lifetime of the spicule-type event is 120 s and oscillations show a period of 60 s. |
| 18 | 08:16:52 | 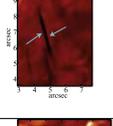 | Spicule-type event present in blue wings at -774 and -1032 mÅ from the line centre. An "M"-shaped swaying motion is observed in the POS, hinting towards some oscillatory behaviour. Lifetime of the spicule-type event is 40 s. |
| 19 | 07:36:10 | 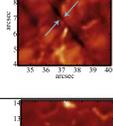 | Spicule-type event present in blue wings at -774 and -1032 mÅ from the line centre. An "M"-shaped swaying motion is observed in the POS, hinting towards some oscillatory behaviour. Lifetime of the event is 70 s and period is 60 s. |
| 20 | 08:03:09 | 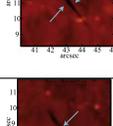 | Spicule-type event present in blue wings at ± 774, and ± 1032 mÅ, from the line centre. An "M"-shaped swaying motion is observed in the POS, hinting towards some oscillatory behaviour. Lifetime of the event is 80 s and period is 70 s. |
| 21 | 08:06:38 | 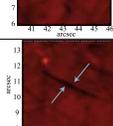 | Spicule-type event present in red wings at +774, and +1032 mÅ, from the line centre. Lifetime and period of swaying motion is 70 s. |
| 22 | 07:46:25 | 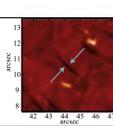 | Spicule-type event present in blue wings and red wing at ± 774, and ± 1032 mÅ. The lifetime of the POS transverse motion is 80 s, which equals the lifetime of the event. However the signature of the event in the red wing lasts for only 1 time-step. |
| 23 | 07:43:17 | 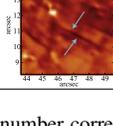 | Blue spicule-type wing event observed at -774 mÅ. Lifetime is 80 s with "W"-shaped swaying motion with a period of 70 s. |
| 24 | 07:46:25 | 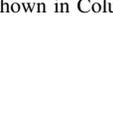 | Blue wing spicule-type event at -774 and -1032 mÅ. Period and lifetime of the swaying motion are nearly same at 80 s. |

**Note.** Column (1) shows a unique number corresponding to the events. Column (2) shows the time corresponding to the first appearance of the event. Column (3) shows the Hα image. All images shown in Column (3) are obtained at +774 mÅ or −774 mÅ. Column (4) describes the events in detail.

Table A4
Table Showing Spicules Selected from 2014 June 07 Data Set

| Case no | Time (hh:mm:ss UT) | Hα image | Comment |
|---|---|---|---|
| 25 | 08:02:57 | | Event present in blue wing at -774 and -1032 mÅ. Period of the swaying motion is similar to the lifetime of the spicule-type event at 100 s. |
| 26 | 07:34:13 | | Event present in blue wings at -774 and -1032 mÅ. Lifetime of the spicule-type event is 100 s, and the period of the swaying motion is nearly 30 s. |
| 27 | 07:50:24 | | Spicule-type event present in blue wings and red wing at ± 774, and ± 1032 mÅ. The "M"-shaped POS motion shows a frequency of 40 s. The lifetime of the event is 50 s. |
| 28 | 07:36:06 | | Spicule-type event present in red wing at +774, and + 1032 mÅ. The "W"-shaped POS motion shows a frequency of 40 s. The lifetime of the event is 45 s. |
| 29 | 08:15:12 | | Spicule-type event present in red wing at +774, and + 1032 mÅ. The "W"-shaped POS motion shows a frequency of 30 s. The lifetime of the event is 45 s. |
| 30 | 08:15:12 | | Spicule-type event present in red wing at +774, and + 1032 mÅ. Swaying motion and lifetime is approximately the same at 30 s. |

## Appendix B
## POS Tracking

Figures B1–B4 show the POS transverse motion corresponding to swaying motion of the spicule-type events. The POS motion is computed using time–distance plots generated along the length of each spicule-type event in image plane, usually with 10 or 20 slits, which are equidistant. The number of slit positions is based on the length of the spicule. These time–distance plots are computed at ±774 mÅ, unless specified. The evolution of POS is traced for spicule-type events, which changes with temporal evolution and is often observed by a distinct "S" shape in dark shading against a light background in the images shown in Figures B1–B4 . The POS motion further represents a change in amplitude and period of motion. Some of this motion appears to be harmonic in nature, where the amplitude and period decrease. Due to the nature of the events, we use one oscillation as the period of the POS motion.

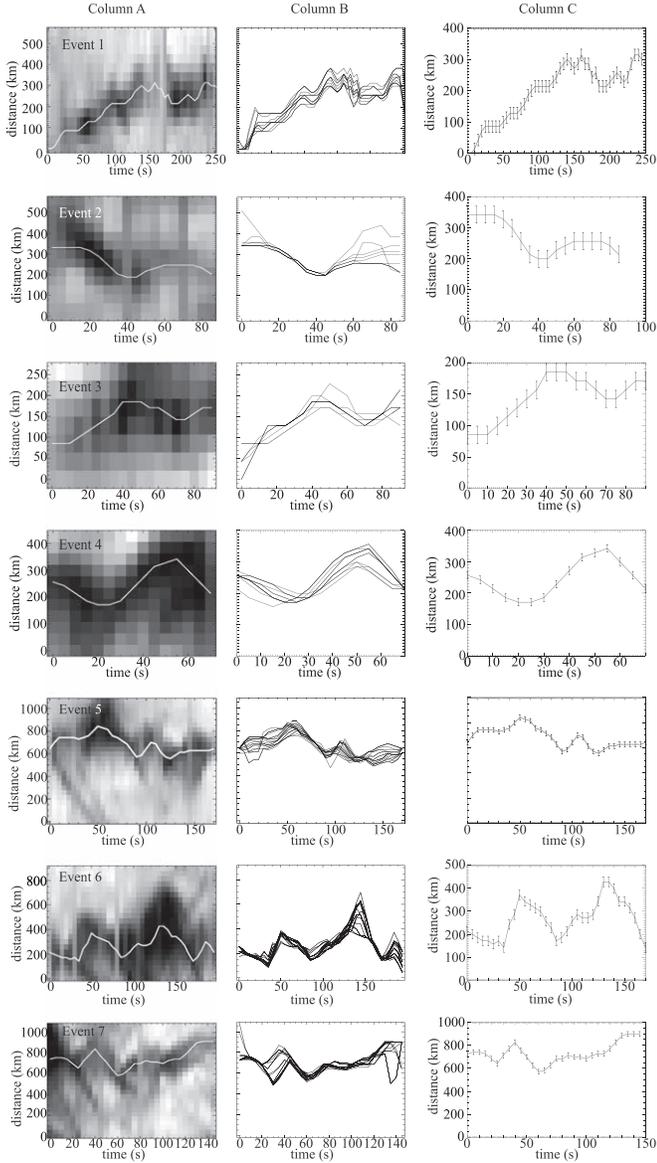

**Figure B1.** Column A shows the best time–distance plot corresponding to the POS transverse motion for event 1–7 from Table A1. Column B shows all the traced POS transverse motion along the length of the spicule-type events obtained at 10 slit positions in the middle of the spicule-type event. Column C shows the track corresponding to the time–distance plot corresponding to the POS transverse motion represented in Column A, with standard error bars overplotted.

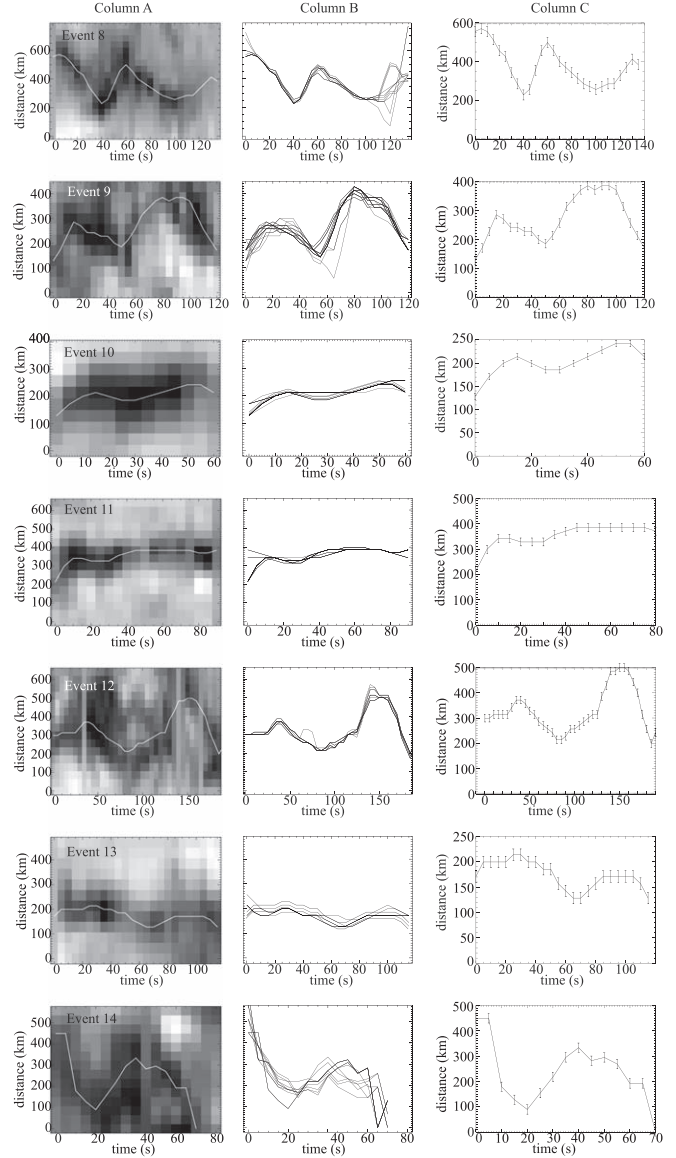

**Figure B2.** Same as Figure B1, but for spicule-type events 8–14 from Tables A1 and A2.

The Astrophysical Journal, 921:30 (24pp), 2021 November 1    Shetye et al.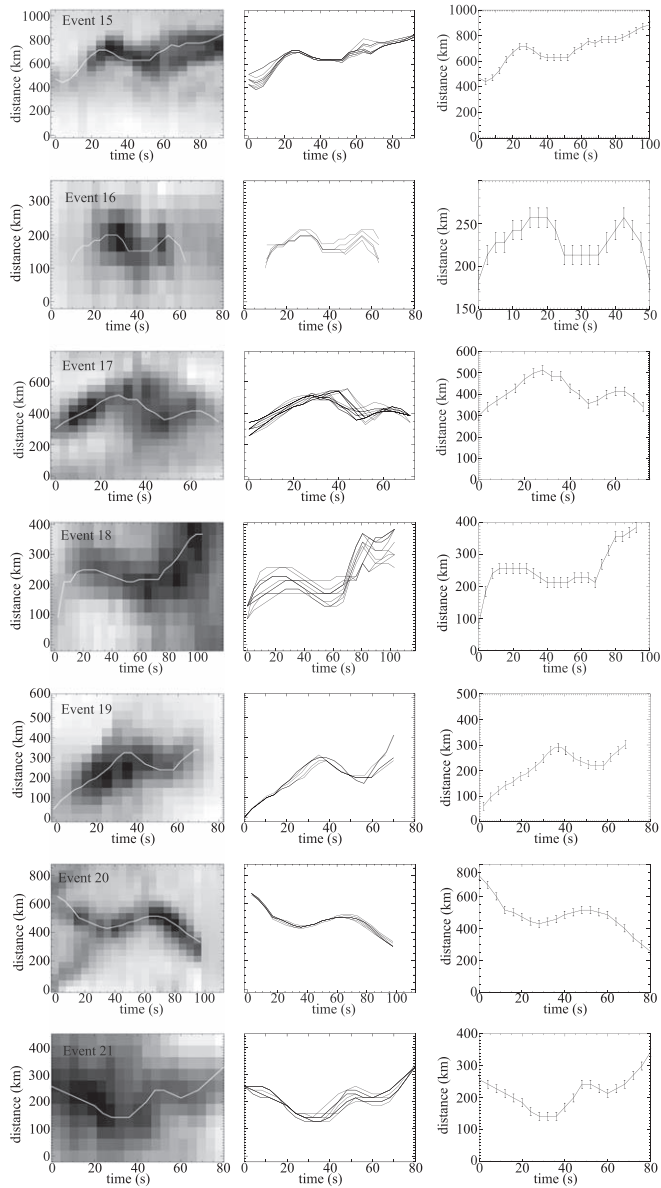

**Figure B3.** Same as Figure B1, but for spicule-type events 17–23 from Table A3.

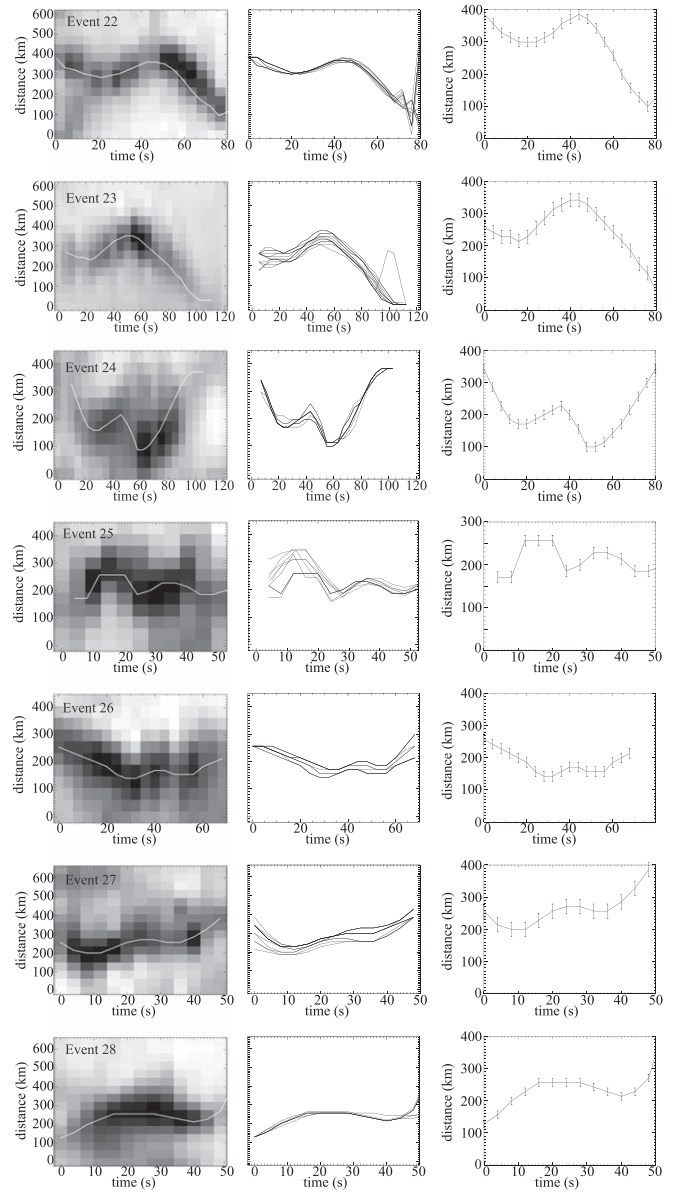

**Figure B4.** Same as Figure B1, but for spicule-type events 24–30 from Table A4.

## ORCID iDs

Juie Shetye ⓘ https://orcid.org/0000-0002-4188-7010
Erwin Verwichte ⓘ https://orcid.org/0000-0002-1723-1468
Marco Stangalini ⓘ https://orcid.org/0000-0002-5365-7546

## References

Antolin, P., Schmit, D., Pereira, T. M. D., De Pontieu, B., & De Moortel, I. 2018, ApJ, 856, 44
Beckers, J. M. 1968, SoPh, 3, 367
Beckers, J. M. 1972, ARA&A, 10, 73
Cavallini, F. 2006, SoPh, 236, 415
De Pontieu, B. 1999, A&A, 347, 696
De Pontieu, B., Carlsson, M., Rouppe van der Voort, L. H. M., et al. 2012, ApJL, 752, L12
De Pontieu, B., Erdélyi, R., & James, S. P. 2004, Natur, 430, 536
De Pontieu, B., & Haerendel, G. 1998, A&A, 338, 729
De Pontieu, B., Hansteen, V. H., Rouppe van der Voort, L., van Noort, M., & Carlsson, M. 2007a, ApJ, 655, 624
De Pontieu, B., McIntosh, S., Hansteen, V. H., et al. 2007b, PASJ, 59, 655
De Pontieu, B., Rouppe van der Voort, L., McIntosh, S. W., et al. 2014a, Sci, 346, D315
De Pontieu, B., Title, A. M., Lemen, J. R., et al. 2014b, SoPh, 289, 2733
Dunn, R. B. 1969, S&T, 38, 368
Fleck, B., & Schmitz, F. 1991, A&A, 250, 235
Haerendel, G. 1992, Natur, 360, 241
Hansteen, V. H. 1997, in in ESA Special Publication, Vol. 404, 5th SOHO Workshop: The Corona and Solar Wind Near Minimum Activity, ed. A. Wilson (Paris: European Space Agency), 45
He, J., Marsch, E., Tu, C., & Tian, H. 2009a, ApJL, 705, L217
He, J.-S., Tu, C.-Y., Marsch, E., et al. 2009b, A&A, 497, 525
Henriques, V. M. J., Kuridze, D., Mathioudakis, M., & Keenan, F. P. 2016, ApJ, 820, 124
Hollweg, J. V., Jackson, S., & Galloway, D. 1982, SoPh, 75, 35
James, S. P., Erdélyi, R., & De Pontieu, B. 2003, A&A, 406, 715
Judge, P. G., Reardon, K., & Cauzzi, G. 2012, ApJL, 755, L11
Judge, P. G., Tritschler, A., & Chye Low, B. 2011, ApJL, 730, L4
Kukhianidze, V., Zaqarashvili, T. V., & Khutsishvili, E. 2006, A&A, 449, L35
Kuridze, D., Henriques, V., Mathioudakis, M., et al. 2015, ApJ, 802, 26
Kuridze, D., Morton, R. J., Erdélyi, R., et al. 2012, ApJ, 750, 51
Kuridze, D., Verth, G., Mathioudakis, M., et al. 2013, ApJ, 779, 82
Langangen, Ø., De Pontieu, B., Carlsson, M., et al. 2008, ApJL, 679, L16723